\documentclass[sn-mathphys-num]{sn-jnl}

\usepackage{graphicx}%
\usepackage{multirow}%
\usepackage{amsmath,amssymb,amsfonts}%
\usepackage{amsthm}%
\usepackage{mathrsfs}%
\usepackage[title]{appendix}%
\usepackage{xcolor}%
\usepackage{textcomp}%
\usepackage{manyfoot}%
\usepackage{booktabs}%
\usepackage{algorithm}%
\usepackage{algorithmicx}%
\usepackage{algpseudocode}%
\usepackage{listings}%

\theoremstyle{thmstyleone}%
\theoremstyle{thmstyletwo}%

\theoremstyle{thmstylethree}%

\usepackage{amsthm,amsmath}
\usepackage[utf8]{inputenc} 
\usepackage{fleqn}
\usepackage{epstopdf}
\usepackage{float}
\usepackage{color}
\usepackage{psfrag}
\usepackage{bigints}
\usepackage{relsize}
\usepackage{amssymb}
\usepackage{txfonts}
\usepackage{mathdots}
\usepackage[classicReIm]{kpfonts}
\usepackage{booktabs}
\usepackage{tikz}
\usetikzlibrary{positioning,decorations,arrows,patterns,shapes,backgrounds}

\usepackage{fancybox}
\usepackage{subcaption}

\usepackage{bm}
\usepackage{array}
\usepackage{amsfonts}
\usepackage{algorithm}
\usepackage{algpseudocode}
\usepackage{pdfcomment}

\newcommand{\xiv}{\mbox {\boldmath $\xi$}}

\newcommand{\muv}{\mbox {\boldmath $\mu$}}
\newcommand{\Phiv}{\mbox {\boldmath $\Phi$}}

\newcommand{\phiv}{\mbox {\boldmath $\phi$}}
\newcommand{\thetav}{\mbox {\boldmath $\theta$}}

\newcommand{\bs}[1]{\boldsymbol{#1}}

\DeclareMathOperator*{\argmin}{\arg\!\min}

\begin{document}

\title[Machine Learning (ML) based Reduced Order Modelling (ROM) for linear and non-linear solid and structural mechanics]{Machine Learning (ML) based Reduced Order Modelling (ROM) for linear and non-linear solid and structural mechanics}

\author*[1]{\fnm{Mikhael} \sur{Tannous}}\email{mikhael.tannous@ensam.eu}
\author[2]{\fnm{Chady} \sur{Ghnatios}}\email{chady.ghnatios@unf.edu}
\author[3]{\fnm{Eivind} \sur{Fonn}}\email{eivind.fonn@sintef.no}
\author[4,3]{\fnm{Trond} \sur{Kvamsdal}}\email{trond.kvamsdal@ntnu.no}
\author[1]{\fnm{Francisco} \sur{Chinesta}}\email{francisco.chinesta@ensam.eu}

\affil*[1]{\orgdiv{PIMM Lab, UMR CNRS}, \orgname{Arts et Metiers Institute of Technology}, \orgaddress{\street{151 Boulevard de l'Hopital}, \postcode{75013}, \city{Paris}, \country{France}}}

\affil[2]{\orgdiv{Department of Mechanical Engineering}, \orgname{University of North Florida}, \orgaddress{\street{1 UNF drive}, \postcode{32224}, \city{Jacklsonville}, \country{United States}}}

\affil[3]{\orgdiv{Department of Mathematics and Cybernetics}, \orgname{SINTEF Digital}, \orgaddress{\street{153 Kloebuveien}, \postcode{7465}, \city{Trondheim}, \country{Norway}}}

\affil[4]{\orgdiv{Department of Mathematical Sciences}, \orgname{Norwegian University of Science and Technology}, \orgaddress{\street{vei 1 Alfred Getz}, \postcode{7491}, \city{Trondheim}, \country{Norway}}}

\abstract{Multiple model reduction techniques have been proposed to tackle linear and non linear problems. Intrusive model order reduction techniques exhibit high accuracy levels, however, they are rarely used as a standalone industrial tool, because of the required high level knowledge involved in the construction and usage of these techniques. Moreover, the computation time benefit is compromised for highly nonlinear problems. On the other hand, non-intrusive methods often struggle with accuracy in nonlinear cases, typically requiring a large design of experiment and a large number of snapshots achieve a reliable performance. However, generating the stiffness matrix in a non-intrusive approach presents an optimal way to align accuracy with efficiency, allying the advantages of both intrusive and non-intrusive methods.This work introduces a lightly intrusive model order reduction technique that employs machine learning within a Proper Orthogonal Decomposition framework to achieve this alliance. By leveraging outputs from commercial full-order models, this method constructs a reduced-order model that operates effectively without requiring expert user intervention. The proposed technique has the possibility to approximate linear non affine as well as non linear terms. It is showcased for linear and nonlinear structural mechanics problems.}


\maketitle

\section{Introduction}
Reduced order modelling (ROM) or model order reduction (MOR) techniques are proposed as an alternative to finite element (FE)/finite volume (FV)/finite difference (FD) full order modeling (FOM), when running parametric optimization for example, requiring the computation of a large number of solutions in a defined parametric interval, and to enable digital twins~\cite{pgd,Quarteroni2016,Rasheed2020dtv,Ghnatios2022}. \\

The present investigation has been initiated in order to enable predictive digital twins for structural health monitoring. Herein, we adopt the following definition of a digital twin~\cite{Rasheed2020dtv}:
"A digital twin is defined as a virtual representation of a physical asset, or a process enabled through data and simulators for real-time prediction, optimization, monitoring, control, and decision-making."\\

To enable predictive digital twins, we advocate the use of hybrid analysis and modelling (HAM) that combines classical physic-based methods (PBM) accelerated by means of ROM together with data-driven methods (DDM) based on sensor measurement analysed by use of machine learning (ML)~\cite{San2021haa,Pawar2021pgm,Pawar2021mfw,Blakseth2022dnn,Blakseth2022cpb,Sorbo2024eem, Stadtmann2023dti}. \\

ROM for structural problems were initially developed using a Galerkin projection procedure with a so-called offline pre-computed reduced order basis that is handling affine problems very efficiently~\cite{pgd,Quarteroni2016}. However, many problems related to structural health monitoring are not affine (e.g., variation of material thickness due to corrosion or nonlinearities due to damages). Some non-affine problems can approximately be made affine by Taylor-expansions (see e.g.,~\cite{Fonn2019fdc}), but this is not generally possible for non-linear problems. To mitigate the challenge with Kolmogorov n-width barrier for non-linear problems many developments has recently been proposed~\cite{BARNETT2023112420}. To alleviate the reduction in MOR's benefits when tackling non-linear problems techniques like approximating the nonlinear term~\cite{poddeim,CHEN2021110545}, or the use of hyperreduction techniques~\cite{hyperreductionfarhat2015}, has been developed. Furthermore, to address the issue of the predefined reduced basis, a stochastic variation along the Grassmann manifold has been proposed \cite{farhatuq0,farhatuq1,mynpmpgd}.\\

The recent development in reduced order modeling and the rising interest in fast and accurate predictions has also lead to the creation of coupled reduced order basis and machine learning techniques. The results of this coupling are hybrid real-time models, able to reproduce with high fidelity the physical solution \cite{Ghnatios2024,BARNETT2023112420}. The ongoing effort in improving these techniques highlights the scientific and industrial interest. However, when it comes to creating a standalone method, able to produce by itself the reduced order model starting from a user input, these techniques represent high intrusiveness and are not easily coupled or integrated in available full order modeling commercial software packages.\\

On the other hand, fully non-intrusive model order reduction techniques are on the rise \cite{spgd1}, with multiple approaches being implemented in commercial software \cite{chinesta2022admore}. These methods do not interfere with the commercial solvers, but operates on a higher level, where first the snapshots solutions are collected, and second, a surrogate model is constructed in the parametric space. These non-intrusive methods can provide real-time simulation inside the parametric space, but may require an excessive number of snapshots offline to correctly construct the surrogate model. Moreover, in presence of strong nonlinearities, the required number of snapshots can become intractable.\\


Given the growing demand for real-time simulations, and in the context of the previously mentioned problem, intrusive Model Order Reduction (MOR) \cite{rom} techniques offers the capability to provide a digital twin for large-scale problems, without the computational expense associated with solving full order models using classical techniques, while maintaining a high fidelity and stability. On the other hand, non-intrusive reduced order models require a large number of snapshots to achieve a reliable performance, especially in the nonlinear context expected in such applications. Moreover, the user may need to increase considerably the number of snapshots without achieving the same accuracy as the one enabled by an intrusive approach. In this work, we suggest combining the strengths of both techniques to develop a lightly intrusive model. This approach suggests generating the inverted stiffness matrix in the reduced space with high accuracy using machine learning algorithms. The aim is to enable high fidelity modeling with reduced computational cost in a nonlinear context while maintaining the user-friendly integration and operation characteristic of non-intrusive methods.\\

The method presented herein aim to enable robust and efficient ROM for non-affine parametric linear problems as well as non-linear problems of non-abrupt nature. The method is characterized by the following properties:
\begin{enumerate}
    \item Inherit the physical properties of the FOM in a lightly intrusive manner by requiring access to the coefficient matrix, right hand side, and the Dirichlet boundary conditions which is typically enabled by commercial software.
    \item Establish a reduced order system by performing a proper orthogonal decomposition (POD) based on the solution vectors from the snapshots obtained by solving the full order problem (FOM) for certain values in the parameter space. This step is similar to what is performed in classical POD-ROM methods \cite{Quarteroni2016,BERNARDI1997209,podhamdouni}.
    \item Flattening the inverse of the reduced stiffness matrix, which consists of a vector space, a second reduction is performed using the principal component analysis on the resulting matrices snapshot columns.
    \item Computing the coefficients able to reconstruct the flattened matrices for every choice of the problem's parameters, leveraging the second reduced order basis, by the means of machine learning algorithms (artificial neural network or random forest for instance).
\end{enumerate}

The main novelty in the proposed approach lies in steps 3 and 4 above, which allows approximating the non-affine parametric terms in the differential equation using a principal component analysis method to create a reduced basis, and a projection method using machine learning algorithm. However, the reader should note that the machine learning algorithm is applied to approximate the inverse of the reduced stiffness matrix, not the solution fields themselves. Unlike approaches such as in \cite{CFD22}, where more than 500 snapshots were needed to accurately predict solution fields using a machine learning-POD coupled method, our lightly intrusive technique focuses on approximating the inverse of the reduced stiffness matrix. This shift in focus is motivated by the need for a generalizable and efficient method to handle non-linear problems, where directly predicting solution fields remains challenging, particularly with a limited number of snapshots, as is the aim in this study. By targeting the inverse of the stiffness matrix, the proposed method ensures broad applicability across various scenarios, minimizing the need for extensive data or complex model adjustments.\\

This article will first revisit the model order reduction techniques in a parametric space, for both intrusive and non-intrusive approaches, in section \ref{morsecall}. In section \ref{problemformulationsec}, we review the elastostatics problem, the selected application in this work. Later on, a simple application is illustrated in section \ref{sec:linearparam}, to showcase the step by step construction of the reduced order model for linear parametric problems. The method is extended to non-linear problems in section \ref{sec:NLGEO}, and for more complex parts in section \ref{sec:airfoil}. Finally, the conclusions and perspectives for future works and improvements are illustrated in section \ref{finalsonclusionssec}. Furthermore, appendix \ref{Abaqus} provides valuable insight into the technical integration of the method within a commercial finite element analysis (FEA) software, specifically Abaqus. This section elucidates the process in a manner that is not only detailed but also generalizable, ensuring that the reader can comprehend the integration independently of the specific application example provided.

\section{On the reduction of parametric fields}\label{morsecall}

For the sake of simplicity, a two-parameter, $\muv=(\gamma,\eta)\in \mathcal{P}$, problem is considered, whose high fidelity solution, assumed for the sake of simplicity to be scalar, is given by $u(\bs x,t,\gamma,\eta)$. We assume that we know in advance $\mathtt P$ solutions related to $\mathtt P$ different choices of both parameters $\gamma$ and $\eta$. In other words, $u(\bs x,t,\gamma_i,\eta_i)\equiv u^i(\bs x,t)$, $i=1, \ldots, \mathtt P$.\\

These known solutions are stored in the form of a matrix, whose columns contain the solution evaluated at each node (rows) at a given time step. The discretized, matrix form of the solution $u(\bs x,t,\gamma_i,\eta_i)\equiv u^i(\bs x,t)$ will hereafter be denoted by $\bm U^i$, with the $(r,s)$-component $u_{rs}^i$ referring to the solution at node $\bs x_r$ at time $t_s$, for the parameter values $(\gamma_i,\eta_i)$, i.e. $u_{rs}^i$ represents $u(\bs x_r,t_s,\gamma_i,\eta_i)$.\\

Finding a parametric regression for each component or degree of freedom of the discrete solution leads to $\bm U(\gamma,\eta)$. This approach, however, becomes in general too expensive and loses smoothness, because the regression's approximation error in the different components remains almost uncorrelated.\\

For this reason, instead of working with the discrete solution itself, reduced space and time bases are extracted by employing the Singular Value Decomposition (SVD) from each solution $\bm U^i$. 
Each matrix $\mathbb U^i$ can be approximated by using the truncated (reduced) bases, that is, there exists a $( K_i \times L_i  )$  $\mathbb D^i$ matrix, approximating $\bm U^i$:
\begin{equation}
\bm U^i \approx \bm X^i \bm D^i \bm T^i,
\end{equation}

where $\bm X^i = (\bm x_1^i \ ... \ \bm x^i_{\mathtt K_i} )$ and $\bm T^{i^T} = (\bm t_1^i \ ... \ \bm t^i_{\mathtt L_i})$.

Matrix $\bm D^i$ could be diagonal when considering $K_i = L_i = R$ and the associated $ R$ highest singular values of $\bm U^i$. 

However, an even better approximation can be performed by assuming $\bm D^i$ non-diagonal and computing it from the minimisation problem
\begin{equation}
\bm D^i = \mathtt{min}_{\bm D^\ast} \| \bm U^i - \bm X^i \bm D^\ast \bm T^i \|,
\end{equation}

with $\bm D^\ast$, $ K_i \times  L_i$ non-diagonal matrices.

The main issue related to the just described procedure is the fact that the reduced basis changes within the parametric space. In order to obtain a common reduced basis, valid for any choice of the parametric solution $\bm U(\mu, \eta)$, we consider the concatenated basis
$$
\tilde{\bm X} = \left [ \bm X^1, \dots , \bm X^{P}\right ],
$$
and
$$
\tilde{\bm T}^T = \left [ \bm T^{1}, \dots , \bm T^{{P}} \right ],
$$
whose SVD decompositions allow extracting the bases $\bm X$ and $\bm T$, composed by $K$ and $L$ modes respectively, enabling the approximation of all the discrete solutions $\bm U^i$, by solving the minimization problem 
\begin{equation}
\bm S^i = \mathtt{min}_{\bm S^\ast} \| \bm U^i - \bm X \bm S^\ast \bm T \|.
\end{equation}

It is assumed now that $\bm S^i$ is known. In other words, we know each component for each sampling $i$, $i=1, ... , P$, of the parameters, $(\gamma_i,\eta_i)$. Therefore,  $S^i_{pq} \equiv S_{pq}(\gamma_i,\eta_i)$, a general parametric expression $S_{pq}(\gamma,\eta)$ is searched, that is, a regression of the scalar components of $\bm S$.

Appendix \ref{pgdregression} provides deeper insight into regularized regression based on PGD.

\section{Reduced elastostatics}\label{problemformulationsec}

The general discretized form of the elasto-statics problem is written:
\begin{equation}\label{dy}
\bm A \bm u = \bm f,
\end{equation}
where the stiffness matrix $\bm A$ and the nodal displacement and force vectors, respectively $\bm u$ and $\bm f$, are of size, respectively, $N\times N$, $N\times 1$ and $N\times 1$.

Matrix $\bm A$ is quite particular because the columns of $\bm A$ have a very  precise physical meaning. By noting $\bm A_{:,c}$ the vector representing the $c$-column of $\bm A$, that vector corresponds to the forces associated with the nodal displacement 
$$
\bm u_c^T=\left (u_1= 0, \dots, u_{c-1}=0, u_c = 1, u_{c+1}=0, \dots , u_N=0 \right ).
$$

Thus, the stiffness matrix $\bm A$ could be viewed as a sequence of forces, representing different nodal unit displacements, the ones composing the canonical basis of $\mathbb R^N$.

{\ } \\
{\em Remark 1.} In the case of elastodynamics the mass matrix will represent the nodal inertia forces. \\

{\em Remark 2.} The non-separability of $\bm A$ (and also the mass matrix in elastodynamics) derives from the fact of having a $N$-rank  matrix affecting the canonical basis, a fact that avoid the existence of a reduced basis for the resulting forces, that is, of the columns of $\bm A$.The diagonal-band structure of mass and stiffness matrices reflects that non-separable structure.

\subsection{Reduced formulation}

In general when forces and displacements are experiencing noticeable correlations, both can be expressed into subspaces of $\mathbb R^N$ of dimension $n \ll N$. 

If we note by $\bm u^i$ the displacement related to the $i^{th}$-parameters choice $\bm \mu_i$, the collected solution snapshots $(\bm u^1, \dots , \bm u^\mathtt P)$, are assembled into matrix $\bf S$. The application of the SVD allows selecting the most relevant orthogonal eigenvectors (modes) that can be grouped into the reduced basis matrix $\bm V$, enabling the global to reduced map:
\begin{equation}\label{mn2}
\bm u \approx \bm V \mathbf \xiv,
\end{equation}
where $\xiv$ is the reduced displacement (or reduced degrees of freedom) vector.

The number of retained modes (highest singular values obtained in the SVD) defines the dimension $n$ of the reduced problem, and consequently the dimension of the displacement and forces subspace, i.e. $\xiv \in \mathbb R^n$.

By introducing the global-to-reduced bases mapping (\ref{mn2}) into the equilibrium equation (\ref{dy}) it results
\begin{equation}
\bm A \bm V \xiv = \bm f,
\end{equation}
that projected into the subspace $\bm V$ leads to the reduced equilibrium formulation
\begin{equation}\label{redelastostatic}
\bm V^T \bm A \bm V \xiv = \bm V^T \bm f,
\end{equation}
that can be rewritten as
\begin{equation}\label{red}
\bm A_r \xiv = \mathbf f_r.
\end{equation}

Since we typically have $n \ll N$, manipulating the reduced equilibrium (\ref{red}) is computationally advantageous with respect of using the global formulation (\ref{dy}).

The same rationale that the one used previously applies on the reduced formulation, and expresses the non-separability of $\bm A_r$.

The preceding section provided an overview of the state-of-the-art application of Proper Orthogonal Decomposition within the context of this research. In the following section, the formulation will be adapted to align with the objectives of this study, specifically by employing Machine Learning (ML) techniques to generate the matrix $\bm{A}_r$, or more precisely, its inverse.

\subsection{Parametric reduced elasto-statics}\label{sec:redelas}

As just commented, the different columns of the reduced stiffness matrix $\bm A_r$ do not accept further reduction, because it has full rank. Thus, if we note the stiffness matrix related to the $i^{th}$-parameters choice as $\bm A_r^i$, $i=1, \dots , \mathtt P$, only correlations in the parametric space are expected. Since, the inverse of the stiffness matrix is utilized for computing quantities of interest, like the displacement vector, we will develop this method such a way machine learning helps in predicting $\mathbf{A}^{-1}$ instead of $\mathbf{A}$. Notably, employing machine learning predictions introduces a certain degree of error in estimating the stiffness matrix. While these errors might be minor individually, they can get magnified during the inverse operation depending on the conditioning of the equation system, potentially resulting in increased inaccuracies in the displacement vector computation.

We proceed by reshaping the inverse of the reduced stiffness matrices into a vector form:
\begin{equation}\label{kvect}
\bm {{A_r^i}}^{-1} \rightarrow \mathcal B^i = \mathtt{Vect}(\bm {{A_r^i}}^{-1}).
\end{equation}
Now, the different inverted stiffness vectors are concatenated
\begin{equation}\label{concatk}
\bm{\mathcal B} = \left ( \mathcal B_1 , \dots , \mathcal B_\mathtt P \right )
\end{equation}

whose SVD decomposition allows obtaining the inverted stiffness vectors reduced orthonormal basis $(\phiv^1, \dots , \phiv^\mathtt R)$. In particular, the projection of $\mathcal B^i$ into the reduced basis allows computing the approximation weights
$$
\mathcal B_i \cdot \phiv^k = \sum \limits_{j=1}^\mathtt R \theta_j^i \ (\phiv^j \cdot \phiv^k), \ \ \ i=1, ... , \mathtt P; \ \ k=1, \dots, \mathtt R,
$$
that taking into account the orthonormality of modes $\phiv^j$ leads to
$$
\mathcal B_i \cdot \phiv^k =  \alpha_k^i,  \ \ i=1, ... , \mathtt P; \ \ k=1, \dots, \mathtt R.
$$

Thus, each inverted reduced stiffness matrix $\mathcal B^i$ is expressed in the common reduced basis by:
\begin{equation}
\label{PhiDecomp}
\mathcal B_i = \sum \limits_{j=1}^\mathtt{R} \theta_j^i \ \phiv^j = \sum \limits_{j=1}^\mathtt{R} \ \phiv^j \theta_j^i = [\phiv^1, \dots, \phiv^R]    \begin{bmatrix}
        \theta_1^i \\
        \vdots \\
        \theta_\mathtt{R}^i
    \end{bmatrix},
\end{equation}

where $i=1, \dots, \mathtt{P}$. Knowing that the different parameter choices $\muv^i$ result in the reduced coordinates $\thetav^i =(\theta_1^i, \dots \theta_\mathtt R^i)$, a regression model can be constructed to predict $\thetav(\muv)$. This prediction facilitates the derivation of the vector form of the associated inverted reduced stiffness, denoted as $\mathcal B_i$, as per Eq. (\ref{PhiDecomp}). Subsequently, this vector representation can be transformed into a matrix form, $\mathcal B_i \rightarrow \bm B_r$. In fact, let's denote $\bm{\Phi}$ the reduction matrix constructed from the reduced basis:
$
\bm{\Phi} = [\phiv^1, \dots, \phiv^R].
$
This means 
\begin{equation}
\mathcal  B_i = \bm{\Phi} \begin{bmatrix}
        \theta_1^i \\
        \vdots \\
        \theta_R^i
    \end{bmatrix}.
\end{equation}

Recalling that ${\bm B}=(\mathcal B_1,\cdots,\mathcal B_\mathtt P)$, with $\mathcal B_i = \mathtt{Vect}({\bm{A}_r^{i}})^{-1} $, we can write:
\begin{equation}
\bm B = \bm{\Phi} \begin{bmatrix}
        \theta_1^1 & \dots & \theta_1^P \\
        \vdots & \ddots & \vdots\\
        \theta_R^1 & \dots & \theta_R^P
    \end{bmatrix} = \bm{\Phi} \bm{\Theta},
\end{equation}
where : 
\begin{equation}
\bm{\Theta} = \begin{bmatrix}
        \theta_1^1 & \dots & \theta_1^P \\
        \vdots & \ddots & \vdots\\
        \theta_R^1 & \dots & \theta_R^P
    \end{bmatrix}.
\end{equation}

\section{Application to a non-affine linear static plate problem}\label{sec:linearparam}

In the following, a linear static parametric example will illustrate the different steps required for the proposed strategy to be set in place, and the main contributions are highlighted.

\subsection{Illustrating the method}

\begin{figure}
    \centering
    \includegraphics[scale=0.4]{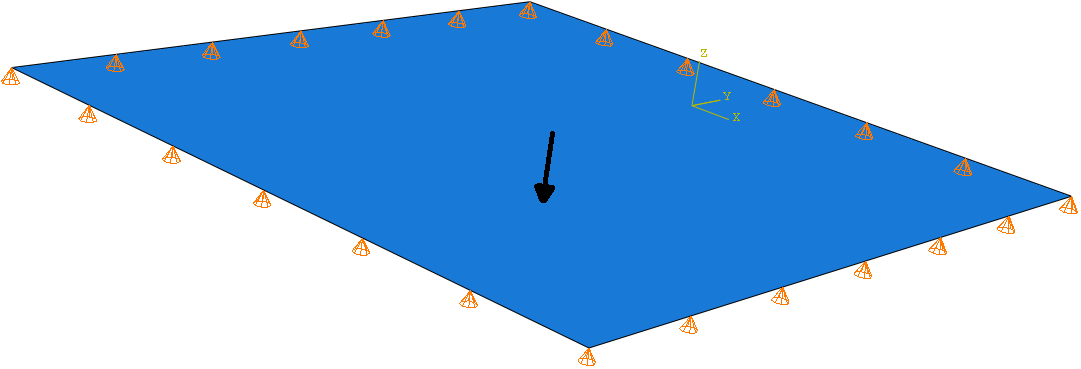}
    \caption{{\em Linear plate problem}: Static point load applied at the centre of a thin metal plate.}
    \label{figthinplate}
\end{figure}

Consider a $1 \times 1$ $m^2$ thin metal plate subjected to a vertical load on its center while its four edges are constrained from moving vertically, as illustrated in Figure \ref{figthinplate}. The plate is discretized with $11 \times 11$ linear shell elements, providing a total of 726 degrees of freedom (DOF).\\

Our analysis involves three varying parameters: the Young's modulus $E$, the Poisson coefficient $\nu$, and the plate thickness $\iota$. Consequently, the stiffness matrix, and subsequently the displacement solution, are functions of these three parameters. Assuming parameter values within the ranges $E \in [100, 300]$ GPa, $\nu \in [0.3, 0.49]$, and $\iota \in [1, 10]$ mm, theoretically, an infinite number of quasi-static problems can be defined within these parameter ranges.\\

Our objective is to compute the displacement vector $\bm u = \bm u(E, \nu, \iota)$ in the online phase without direct access to the stiffness matrix generated by the full order FE software for any combination of $(E, \nu, \iota)$ in the range previously defined. To achieve this we here express the inverse of the stiffness matrix by means of a reduced orthonormal basis $(\phiv^1, \dots, \phiv^\mathtt{R})$ as stated in Eq. (\ref{PhiDecomp}) through a pre-trained machine learning algorithm (ML) that predicts the matrix coefficients in the reduced basis, thus providing ${(\mathbf{A}_r^i})^{-1}$ for a given set of parameters $\mu_i$.\\

The subsequent steps, outlined below, apply the principles discussed in previous sections.\\

First, a basis of solution vectors must be constructed, where each solution vector $\bm u^i$ is associated with specific parameters $\mu^i=(E_i, \nu_i, \iota_i)$. The snapshot size, denoted as $\mathtt{P}$, determines the number of experiments, and in this case, the snapshot size is set to $\mathtt{P}=65= 4^3+1$. The parameters combination associated with each experiment can be chosen using a Latin-Hyper Cube \cite{lhs} or also using the zeros of the Chebyshev quadrature points \cite{cheby}, among other techniques. For the problem under consideration, we used the zeros of the Chebyshev polynomials of fourth order in each of the three parameter dimensions plus the point in the centre of the parameter domain.\\

For each snapshot, a FE solver is utilized to perform the calculation. For each experiment $i$, the associated stiffness matrix $\bm {A}^i$, displacement vector $\bm{u}^i$, and right-hand force vector $\bm{f}^i$ are recorded. Please refer to appendix \ref{Abaqus} for more details on how this is technically achieved in a commercial software. Singular Value Decomposition (SVD) is then applied to the collected solution snapshots $(\bm{u}^1, \dots, \bm{u}^\mathtt{P})$. By retaining the significant modes, a reduction matrix $\bm{V}$ is obtained, which is used to reduce the stiffness matrix $\bm{A}^i$ to $\bm{A}_r^i$, as expressed by:
\begin{equation}
\bm{A}_r^i = \bm{V}^T \bm{A}^i \bm{V}.
\end{equation}
The linear static plate problem is now presented in the form of Eq. (\ref{redelastostatic}). The size of the stiffness matrix $\bm{A}^i$, initially $N \times N = 726 \times 726$, is reduced to $\bm{A}_r^i$ of size $n \times n = 2 \times 2$ for the presented case by retaining the first set of eigenvalues such that the ratio between the largest eigenvalue and the smallest retained one is $1000$.  This matrix is then inverted to $({\bm{A}_r^{i}})^{-1}$ and reshaped into a vector form ${\mathcal B}_i$. The subsequent steps, as described in Section \ref{sec:redelas}, remain unchanged.

%
%

In this example, since $R$ is equal to 2, only two coordinates in the stiffness matrix reduced space are required to generate $\mathcal A^{-1}_i$. Reshaping it to a $n \times n$ matrix form allows solving the elasto-static problem by transforming Eq. (\ref{red}) for a given set of parameters $\mu_i$ using:
\begin{equation}
\xiv^i = {\bm{A}_r^{i}}^{-1} \bm f_r^i,
\end{equation}
where $\bm f_r^i$ obtained by projecting the right-hand side load vector $\bm f^i$ on the reduced basis $\bm V$, using $\bm f_r^i = \bm V^T \bm f^i$, and then deducing the global displacement vector $\bm u$, using Eq.~(\ref{mn2}).\\

A machine learning algorithm must now be selected and trained to predict $\bm{\Theta} = \bm{g}^{\textsc{ML}}(E, \nu, \iota)$. Although Artificial Neural Networks (ANNs) are widely used and have demonstrated the capability to achieve highly accurate predictions on this example, they will not be employed in this research. The primary objective is to select a machine learning technique that requires minimal user intervention or tweaking during training. Given that the method is intended for application across various structure sizes and types, ANNs present the disadvantage of requiring architecture modifications depending on the data type and size. After testing XGBoost \cite{XGBoost}, Support Vector Machines \cite{SVM}, and Random Forest \cite{RFregressor}, the latter emerged as the best performer in terms of accuracy and overall performance. The Random Forest Regressor not only matches the precision of a well-trained neural network for the tested cases but also circumvents the challenges associated with determining the optimal neural network architecture, as its default settings are sufficient to achieve satisfactory accuracy. Moreover, although the snapshot size is 65, only 54 randomly chosen snapshots are used for training the network, while the remaining are used as a validation set. Furthermore, any use case can be generated by selecting an appropriate set of parameters for $\mu=(E, \nu, \iota)$ within the originally specified range, allowing a comparison between the Finite Element Method (FEM) solution and the one deduced using the neural network prediction.

\subsection{Results discussion}

As a first validation step, a comparison between the $\bm{\theta}_j$ outputted by the machine learning prediction and a reference one obtained from the inverted reduced stiffness matrix, is performed. $\bm{\theta}^{ref}_j=\bm{\Theta}_{ref}[j,:] $ with:
\begin{equation}
\mathbf{\Theta}_{ref} = ({\bm{\Phi}^{T}}\bm{\Phi})^{-1}{\bm{\Phi}^{T}} \mathcal B 
\end{equation}

\begin{figure}
\centering
    \includegraphics[width=\textwidth]{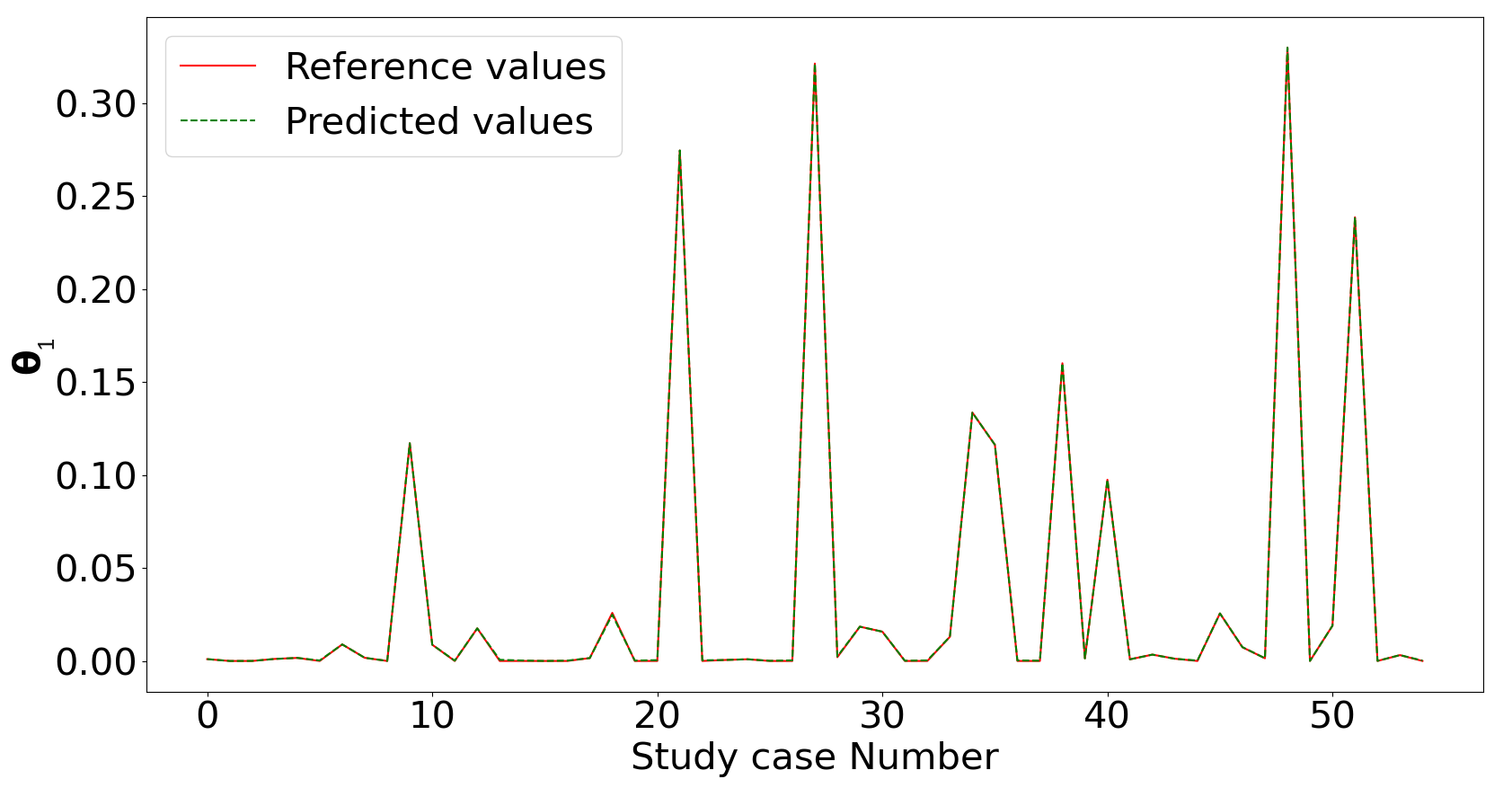}
    \caption{{\em Linear plate problem}: Predicted versus reference $\bm\theta_1$ coefficients on the training data set.}
    \label{fig:phi1tr}
\end{figure}

In analyzing the training set employed to train random forest prediction functions $g^{\textsc{ML}}(E, \nu, \iota)$, we observe from Figure~\ref{fig:phi1tr} that the predicted values remarkably mirror the reference values with exceptional precision. This fidelity holds true across all reduced coordinates of the stiffness matrix, whether on the training or validation datasets. Notably, the variance between the reference coefficients $\bm\theta_j$ ($j=1,\ 2$) and their predicted counterparts is virtually imperceptible to the naked eye, requiring significant magnification to discern.\\ 

\begin{figure}
\centering
    \includegraphics[width=\textwidth]{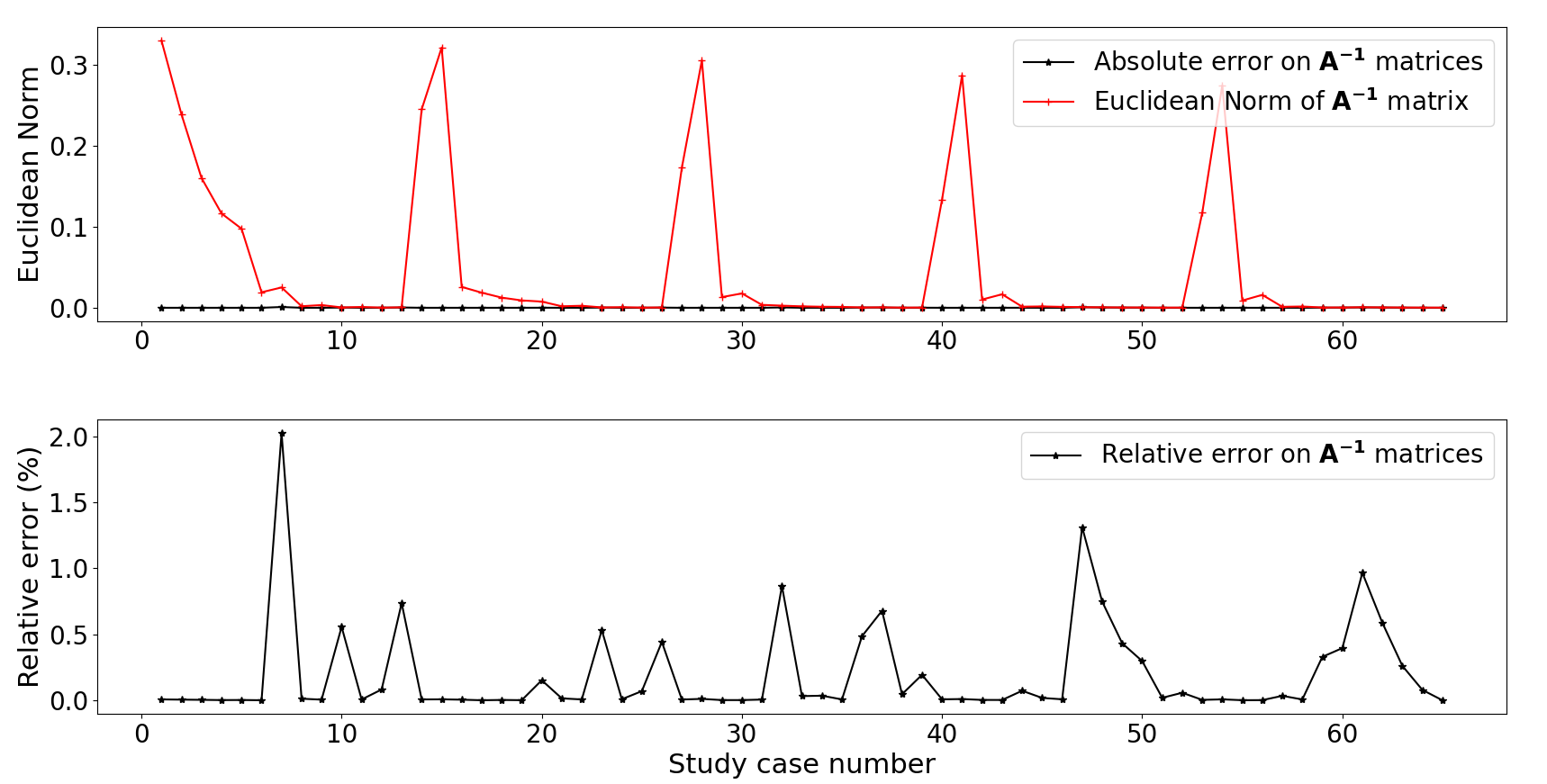}
    \caption{{\em Linear plate problem}: Evaluating precision through a comparison of the Frobenius norm between the reference FEM stiffness matrix and the matrix predicted using machine Learning (ML)- based Reduced Order Modeling (ROM).}
    \label{fig:kcoefferr}
\end{figure}

\begin{figure}
\centering
    \includegraphics[width=0.9\textwidth]{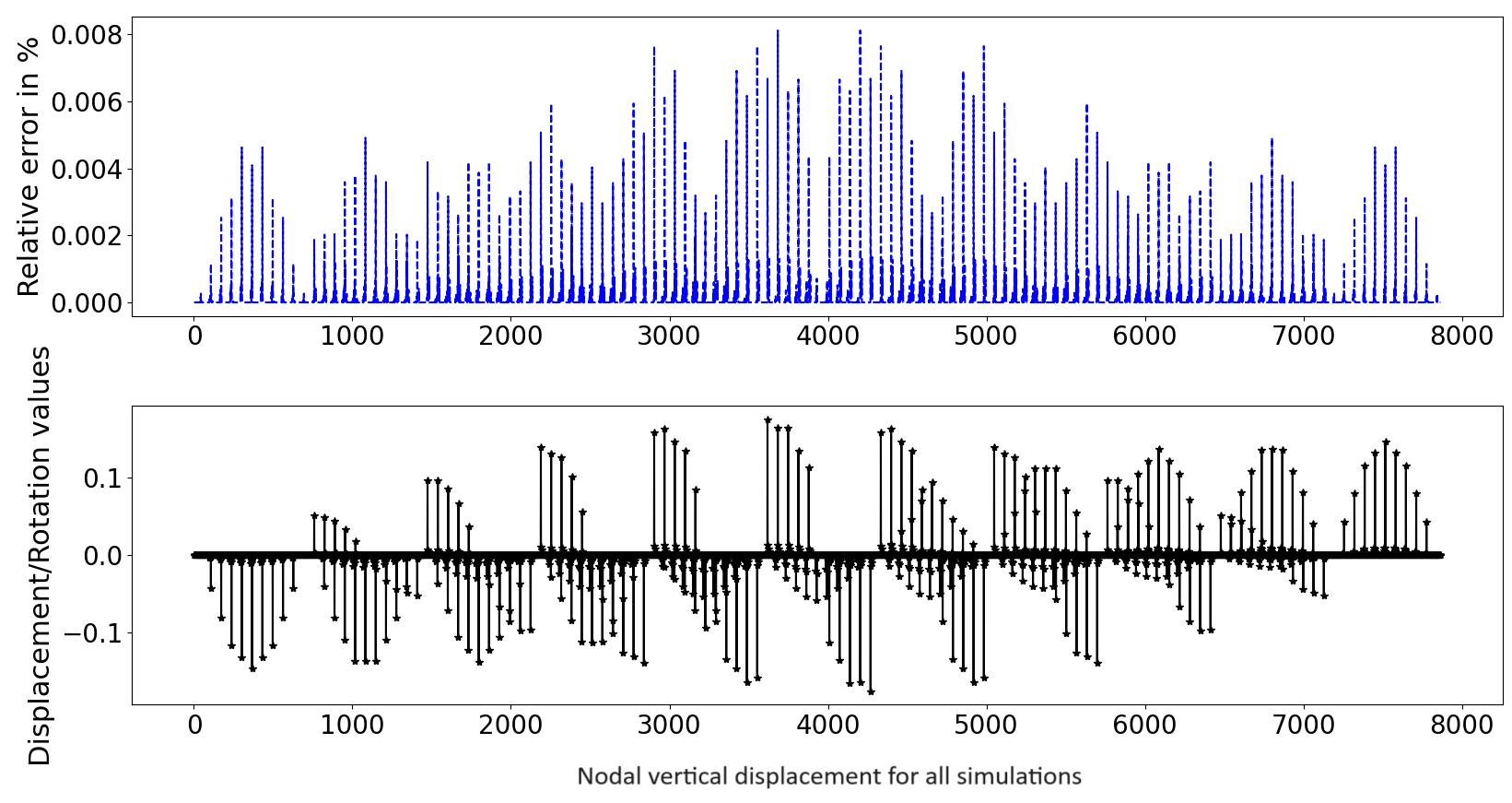}
    \caption{{\em Linear plate problem}: Assessing the prediction error on the vertical displacements}
    \label{fig:u3disp}
\end{figure}

\begin{figure}
\centering
    \includegraphics[width=1\textwidth]{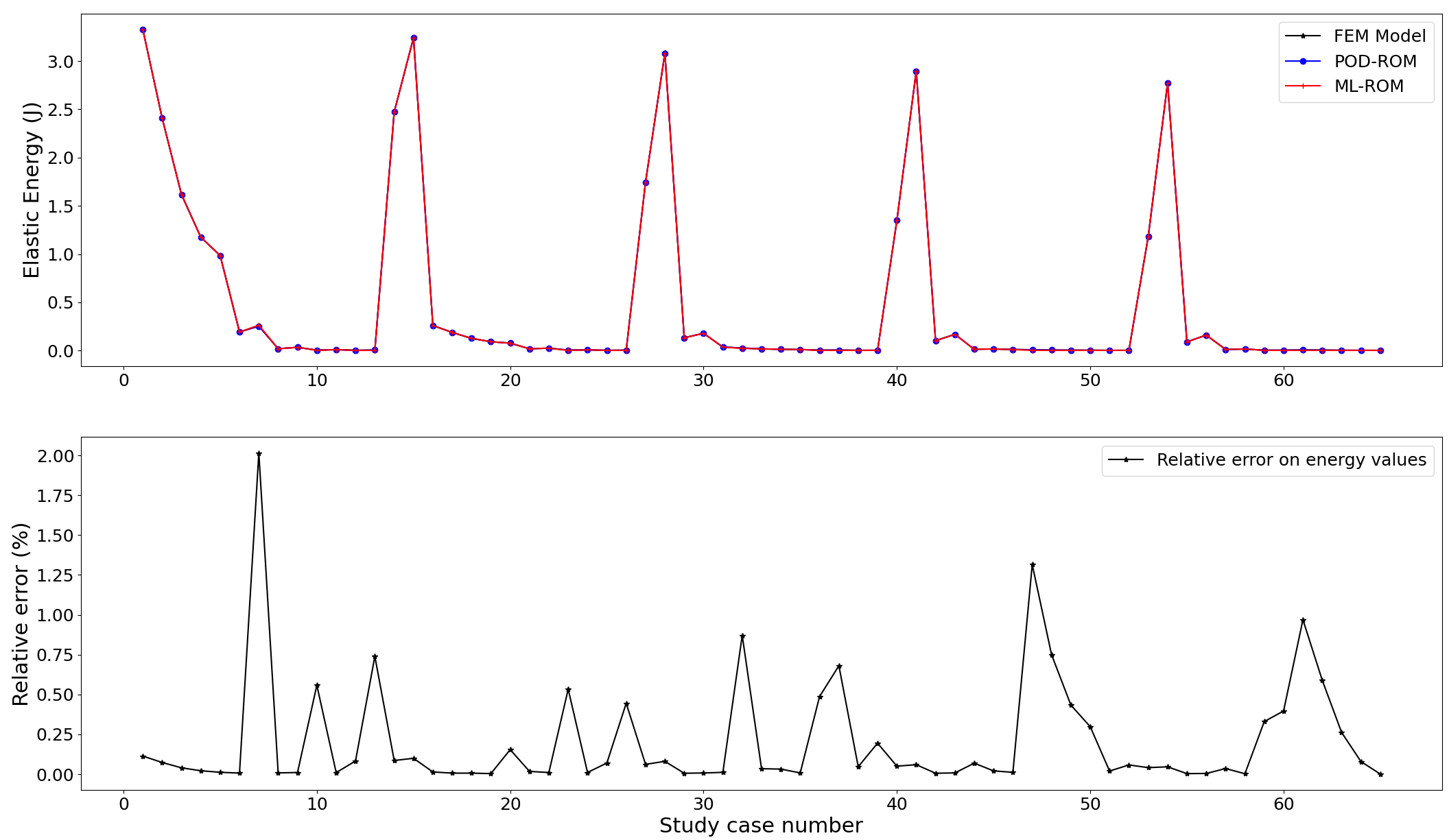}
    \caption{{\em Linear plate problem}: Comparison of the computed elastic energy and corresponding relative error in energy for the 65 snapshot cases}
    \label{fig:ElasticEnergy}
\end{figure}

It's worth noting that the study cases are independent entities. Consequently, the connection between the $\bm\theta_1$ values depicted in Figure~\ref{fig:phi1tr} lacks any inherent physical significance. However, establishing connections among the diverse $\bm\theta_1$ values corresponding to each study case serves the purpose of facilitating visual comparison between the two solutions. This principle extends to Figures~\ref{fig:kcoefferr}, \ref{fig:u3disp}, and \ref{fig:ElasticEnergy}, where such visual aids streamline the comparative analysis process.\\

The $\bm{\Theta}$ have direct impact on the prediction of the stiffness matrix. We can assess this impact by comparing the $\mathcal{A}$ values associated with the reduced order model (ROM) solution to those predicted using the ML-informed ROM. For each study case $i$, we can compute the error $\Delta \mathcal{B}_i$ as the Frobenius norm of the difference between the reference and the predicted $\mathcal{B}_i$ vectors:
\begin{equation}
\Delta \mathcal{B}_i =||\mathcal{\bf B}_i^{\text{reference}}-\mathcal{\bf B}_i^{\text{predicted}}||_{\textrm{F}}.
\end{equation}
Therefore, the relative error on the prediction of $\mathcal{B}_i$ vectors, in percent, is:
\begin{equation}
\Delta \mathcal{B}_i^r = \frac{||\mathcal{\bf B}_i^{\text{reference}}-\mathcal{\bf B}_i^{\text{predicted}}||_{\textrm{F}}}{||\mathcal{\bf B}_i^{\text{reference}}||_{\textrm{F}}}\times 100.
\end{equation}

As Figure~\ref{fig:kcoefferr} shows, the error on the Frobenius norm of the stiffness matrix is low with a relative error not exceeding $2\%$. The maximum relative error corresponds to the lower matrix coefficients.\\

\begin{figure}
\centering
    \includegraphics[width=\textwidth]{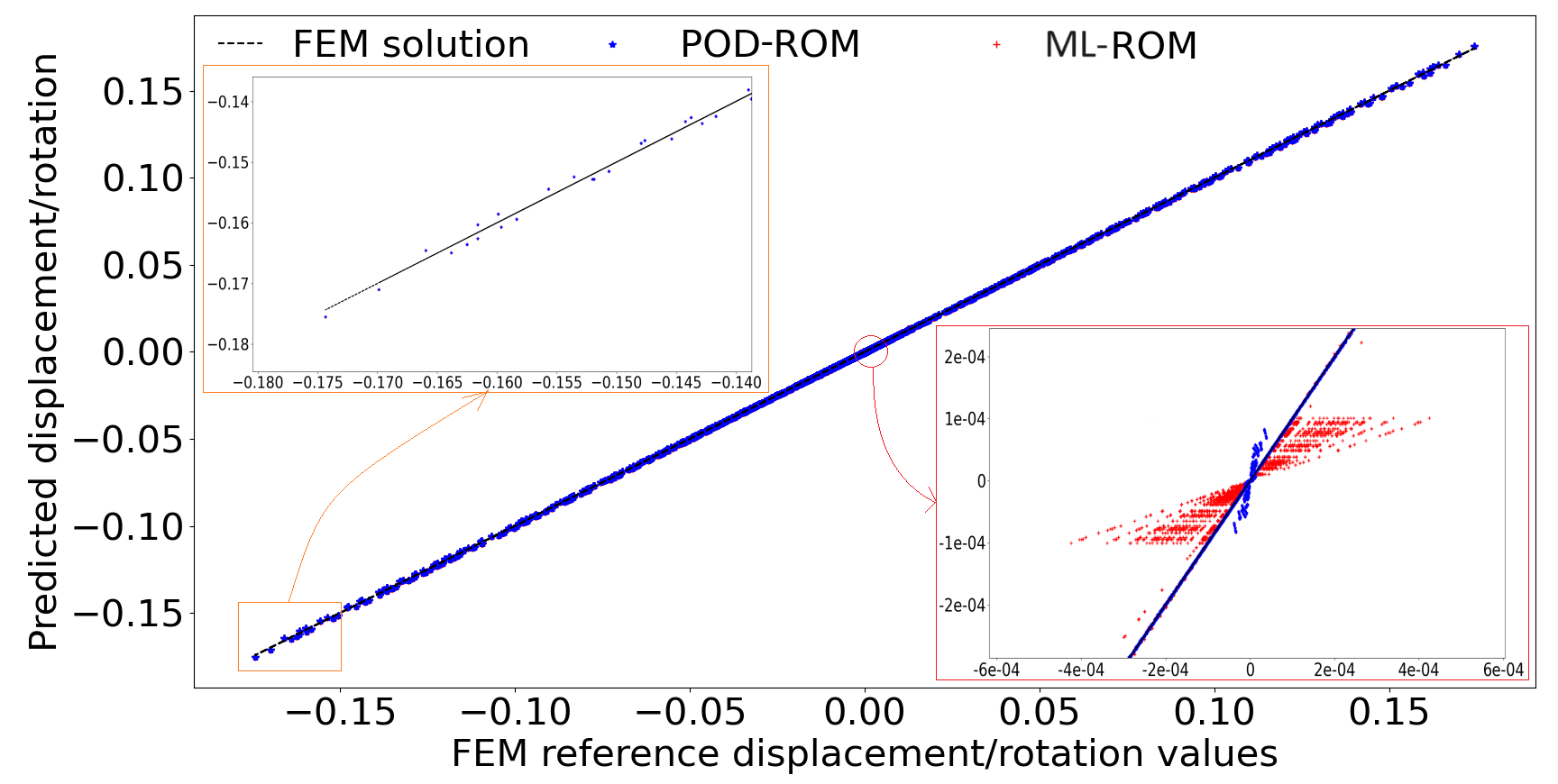}
    \caption{{\em Linear plate problem}: Assessing the precision of the method on displacement and rotation values for all degrees of freedom associated with different training and validation datasets.}
    \label{fig:disp}
\end{figure}

\begin{figure}
\centering
    \includegraphics[width=\textwidth]{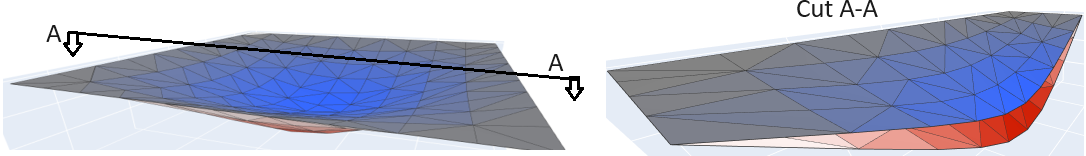}
    \caption{{\em Linear plate problem}: Comparison of the plate deformation, scaled by a factor of $1000$ for sake of better visibility. In blue the results obtained with the Machine Learning (ML) based Reduced Order Model (ROM) and in red the corresponding finite element based Full Order Model (FOM) results.}
    \label{fig:defplate}
\end{figure}

In the comparison above, the machine learning predictions yield satisfactory results. Subsequently, we delve deeper by examining the actual quantities of interest, such as displacements. This examination offers a nuanced understanding of the precision of our proposed method, particularly as we can juxtapose it with the global displacements generated by the FEM solver, and those provided by the POD-ROM. For sake of completeness, we compute displacements for every degree of freedom across all 65 calculations. Given 726 degrees of freedom (DOF) for each set of parameters $\mu_i$, a total of 47190 values are compared, encompassing both validation and training datasets as prediction precision is similar on training and validation datasets. Figure~\ref{fig:disp} plots the predicted versus reference displacements/rotations for all DOFs, with the $y=x$ line representing zero error values. Points further from this axis indicate higher error. Both standard POD-ROM and ML-ROM solutions are displayed. As depicted, displacements obtained using the machine learning-informed ROM closely resemble those obtained using classical POD-ROM, which, in turn, agree with the FEM solution. However, upon closer examination of low displacement values, particularly near the $(0,0)$ center, higher errors are observed in predictions. Nonetheless, for higher displacement values, the ML-ROM closely mirrors its POD-ROM counterpart. Notably, errors in low displacement values, such as those of nodes near boundary conditions or nodes with negligible rotations, bear minimal physical significance. Indeed, analysis of a test case with the highest displacement error reveals that the FEM deflection and the ML-ROM one are indistinguishable unless displacements are magnified by a factor of 1000, as demonstrated by cut A-A in Figure~\ref{fig:defplate}. Therefore, high errors are primarily associated with non-significant displacements and rotations that have minimal impact on the physical deflection of the plate.\\

To quantify the error on the most relevant degrees of freedom, we illustrate the error appearing on the vertical displacements only in Figure~\ref{fig:u3disp}, for both the training and validation sets. In the latter, all $121$ nodes belonging to all $65$ study cases are present and numbered from zero to $7865$. Error assessment is performed by diving the displacement error by the average deflection value. As Figure~\ref{fig:u3disp} shows, the vertical displacements exhibit minimal prediction errors. The maximum error does not exceed $0.008\%$.\\

Another relevant accuracy metric is the elastic energy. We choose to plot the elastic energy associated with each of the 65 calculations. We compare in Figure~\ref{fig:ElasticEnergy} the elastic energy of the FEM solution with the one of the POD-ROM and of the ML-ROM. All three curves match very well, illustrating therefore the precision of the proposed approach.

Moreover, as Figure~\ref{fig:ElasticEnergy} shows, the relative error remains lower than $2\%$ with the larger error associated with the lower energy cases. These conclusions confirm the ones obtained using the displacements and the matrix coefficients.

\subsection{Conclusion}

Based on the aforementioned comparisons and error metrics, it can be concluded that the proposed method is performing well. It enables accurate prediction of the reduced inverse of the stiffness matrix for parameter combinations that were not part of the training set, thereby facilitating precise computation of the displacement fields across all degrees of freedom in an elastostatic problem.
\section{Application to a non-affine  geometrically nonlinear plate problem}\label{sec:NLGEO}

Following the successful application of the method to parametric elastostatic problems, our objective now shifts towards its extension to nonlinear scenarios. Given the substantial variability in nonlinearities, the adaptation of the method to address such cases necessitates a tailored approach contingent upon the specific nature of nonlinearity under consideration. In this paper, we address geometrically nonlinear problems as a preliminary step. The method's success in this endeavor will lay the groundwork for highly nonlinear applications in future works.\\

In a geometrically nonlinear scenario, the stiffness matrix becomes a function of displacement, while the load vector remains a known entity within the problem domain. To illustrate, consider a simple 1D problem, such as the deformation of a spring under the influence of an external force $F$. In static equilibrium, the spring stiffness $K$ can be represented as:

\begin{equation}
K(x) = \frac{dF}{dx}.
\end{equation}

The integration process unfolds iteratively over $N_t$ steps. For each step $i$ within the range $[1,N_t]$, employing a numerical integration method like the trapezoidal rule yields:
\begin{equation}\label{eq:spring_nl}
 F_{i} - F_{i-1} = \frac{ \left( K_{i} +  K_{i-1} \right )}{2} \left( x_{i} -  x_{i-1} \right )    
\end{equation}
with $i=0$ corresponding to initial conditions. This fundamental principle extends seamlessly to 2D or 3D problems, albeit with variations in the integration techniques depending on the Finite Element Method (FEM) solver employed. Nevertheless, the trapezoidal rule remains effective for geometric non-linearity, particularly when the number of integration steps $N_t$ is sufficiently large.\\

Readers should note that the core steps for generating the inverted stiffness matrix remain consistent and that the method itself is generalizable. The detailed discussion on how the the integration scheme is used to solve the geometrically nonlinear problem is necessary because we must compare the Machine Learning-based Reduced Order Model (ML-ROM) solution with both the Proper Orthogonal Decomposition-based Reduced Order Model (POD-ROM) and the reference solution from the Finite Element Analysis (FEA) software. To ensure a fair comparison, it is essential to use the same integration scheme, time step, and other parameters as those used by the commercial software. In the following sections, it will be evident that an average stiffness matrix is generated, rather than one specific to a given time step?a choice dictated by the selected commercial FEA software. If the method is applied independently, users can select their preferred integration scheme and proceed accordingly. The primary difference between linear and nonlinear applications lies in the parameters to be considered when generating the stiffness matrix. For linear problems, material parameters are sufficient, while for nonlinear problems, the displacement at the previous time step must be included, as the stiffness matrix becomes solution-dependent. Appendix \ref{Abaqus} provides a technical explanation on how the method is implemented in Abaqus, presented in a general format that is adaptable to various use cases.\\

In subsequent sections, we proceed to prove the application of this methodology. Section~\ref{sec:platestatic} serves as a starting point, wherein we showcase the method's efficacy through its application to a 2D plate experiencing a geometric nonlinearity induced by an external loading. In this case, the material parameters are well-defined. The primary challenge, as opposed to the linear parametric case detailed in Section~\ref{sec:linearparam}, lies in devising an appropriate integration technique.\\

Later on, Section~\ref{sec:plateparametric} delves into the extension of the methodology to address parameterized problems exhibiting nonlinear geometric behavior. Parametric nonlinear problems consists a significant advancement, presenting further complexities in integration strategies and solution methodologies.

\subsection{Displacement prediction for fixed material parameters}\label{sec:platestatic}

\begin{figure}
\centering
\includegraphics[width=0.7\textwidth]{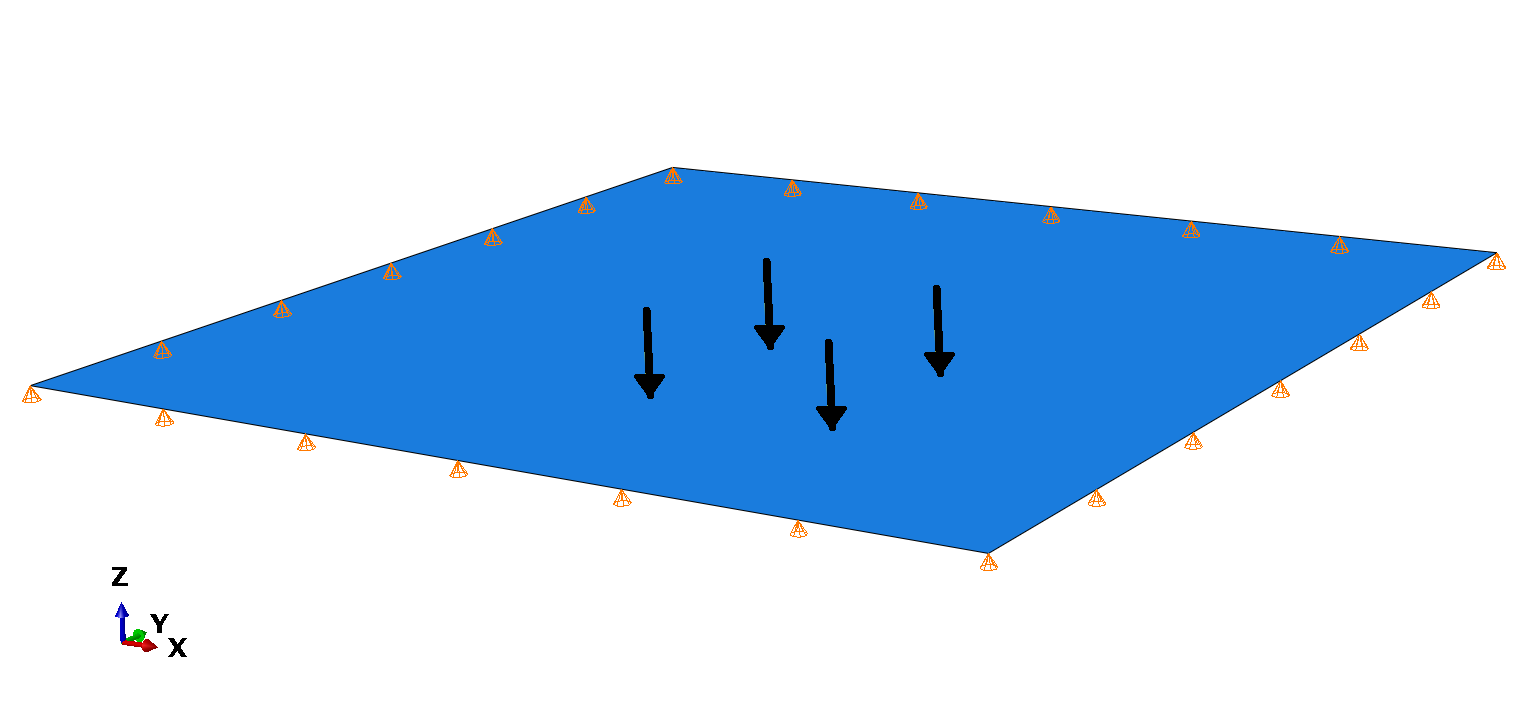}
\caption{{\em Geometrically nonlinear plate problem}: The plate under study along with the force vector consisting of four point-loads.}
\label{fig:NL_Plate}
\end{figure}

In this section, we reexamine the plate introduced in Section~\ref{sec:linearparam}. However, we are now exploring it through a different lens. The geometrical and material parameters are now fixed, providing a controlled environment for our investigation. Specifically, the Young's modulus remains constant at 100 GPa, with a Poisson coefficient fixed at 0.3, and the plate thickness set to 1 mm. The loading configuration entails the application of a force uniformly distributed across four points at the middle of the plate, as depicted in Figure~\ref{fig:NL_Plate}. This force is increasing as a function of the time along the vertical direction, starting from zero and reaching 400 kN at $t=1 s$.\\

Upon reaching equilibrium, the applied load induces a deformed configuration with a maximum deflection of 7.32 cm.\\

Solving the nonlinear geometric problem involves $N_t=118$ time steps executed by the selected commercial FE solver, utilizing the trapezoidal integration method. Throughout each integration step, the stiffness matrix, displacement, and load vectors are recorded. Solving a geometrically nonlinear static problem could be approached differently from one FE solver to the other. However, a widely used approach consists in applying the load gradually starting at $t=0$ with a load free structure and reaching equilibrium at $t_{end} = 1$ with the load fully applied. This one-second time lapse is divided into time steps, with $t_i$ referring to the $i^{th}$ time instant. Since it is a static problem, time is fictitious and $t_i$ could be seen as an iteration step. The time increment size $\delta t_i = t_{i} - t_{i-1}$ may evolve as a function of $i$ depending on the software's solving algorithm. The governing equation for the plate, solved using the trapezoidal rule at $t_i$, generalized from Eq.~(\ref{eq:spring_nl}), reads:

\begin{equation}\label{eq:plate_nl_solve}
 \frac{ \left(\bm{A}_{i} + \bm{A}_{i-1} \right )}{2} \left( \bm{u}_{i} - \bm{u}_{i-1} \right ) = \bm{f}_{i} - \bm{f}_{i-1}
\end{equation}

We denote $\tilde{\bm{A}}_i = \frac{1}{2} \left( \bm{A}_{i} + \bm{A}_{i-1} \right)$ as the average stiffness matrix at time step $t_i$. Each stiffness matrix $\tilde{\bm{A}}_i$ and displacement vector $\bm u_i$ are associated with their corresponding integration step $t_i$, following the methodology of Section~\ref{sec:linearparam}. First of all the solution space is reduced by using a first SVD on the solution snapshots $\bm u_i$, leading to the reduced basis $\bm V$. Using $\bm V$, $\tilde{\bm{A}}_i$ are reduced to $\tilde{\bm{A}_r}_i$, which are in turn inverted and then reshaped into a vector shape $\mathcal{B}_i$. Once the second SVD is used on the collection of matrix snapshots $\mathcal{B}=(\mathcal{B}_1,\cdots,\mathcal{B}_{N_t})$, we obtain a matrix $\bm \Theta$ of size $(5 \times N_t)$ encapsulating the essential information of the reduced order model, facilitating therefore the reconstruction of the displacements' evolution under the imposed loading conditions. The dimensional reduction to a size of 5 is determined by selecting the first set of eigenvalues such that the ratio between the highest eigenvalue and the lowest retained eigenvalue is $1000$.\\

For each time step $t_i$, we dispose five ${\theta}_i$ values necessary to reconstruct $\mathcal{A}_i$, or more precisely, its reduced inverse denoted as vector $\mathcal{A}^{-1}_i$. This vector is reshaped into the corresponding reduced matrix and then used to obtain the reduced coordinates $\xiv_i$. Subsequently, the global degrees of freedom are deduced using $\bm u_i = \bm{V} \xiv_i$. The different reconstruction steps are illustrated in Algorithm \ref{algo1}. Note that $\bm V^T \left( \mathbf{f}_{i} - \mathbf{f}_{i-1} \right)$ can be precomputed in the offline phase if the right hand side $\mathbf{f}_i$ is known.\\

To minimize ambiguity, we propose the following notation for key elements derived from matrix $\bm \Theta$:
\begin{itemize}
    \item $\underline{\bm \theta}_i$: This vector contains the reduced matrix coefficients at time step $i$, enabling the reconstruction of the inverse of the reduced stiffness matrix for that specific time step.
    \item ${\bm \theta}_k$: This vector encompasses all information pertaining to the $k^{th}$ reduced coefficient across various time steps.
\end{itemize}

\begin{algorithm}
\caption{Iteratively Updating $\bm\Theta$ and Solving for $\bm u$}
\label{algo1}
\begin{algorithmic}[1]
\State \textbf{Step} $i = 0$: $\bm u_i=0 \implies \xiv_{0}=0$. Define $\delta t_1$.
\For{$i = 1$ \textbf{to} $N_t$}
\State $\underline{\bm \theta}_i \gets$ function of $(\xiv_{i-1}, \delta t_i)$ \Comment{Update $\bm\Theta$ based on previous state and increments}
\State Deduce ${\bm{A}_r^{-1}}_i$ based on $\underline{\bm \theta}_i$ and $\mathbf{V}$ \Comment{Update the inverse stiffness matrix}
\State $\xiv_i \gets \xiv_{i-1} + {\bm{A}_r^{-1}}_i \bm V^T \left( \mathbf{f}_{i} - \mathbf{f}_{i-1} \right)$ \Comment{Solve for reduced displacements}
\State Deduce $\bm u_i = \bm{V} \xiv_i$ \Comment{Update $\bm u$}
\EndFor
\end{algorithmic}
\end{algorithm}

The machine learning training will enable the prediction of $\underline{\bm \theta}_i$ at each time step $t_i$ based on the reduced displacements obtained from the previous time step, the current time step $t_i$ and the time step increment $\delta t_i$ (which defines the next time step $t_{i+1}$). For the training, we consider $85 \%$ of all time steps as part of the training data set and the remaining time steps serve as the validation data set.\\ 

\begin{figure}
\centering
\includegraphics[width=\textwidth]{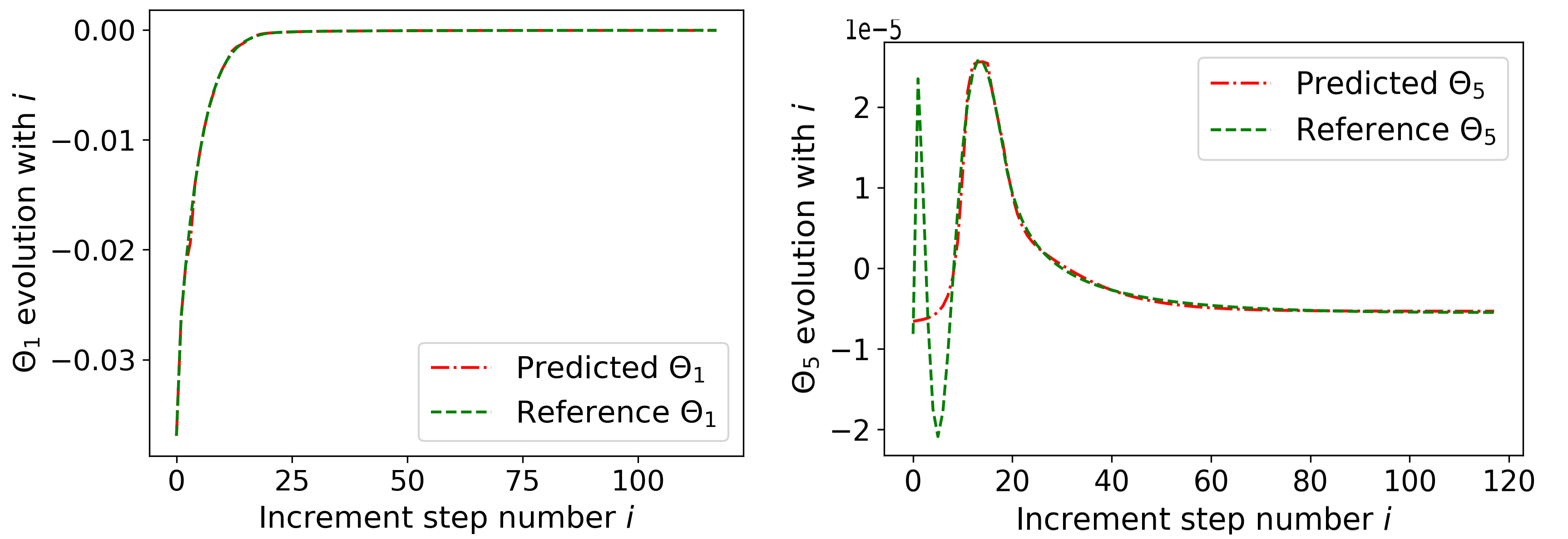}
\caption{{\em Geometrically nonlinear plate problem}: Evolution of $\bm\theta_1$ and  $\bm\theta_5$ as a function of the increment step number.}
\label{fig:NLGeo_Phi_1_5}
\end{figure}

\begin{figure}
\centering
\includegraphics[width=\textwidth]{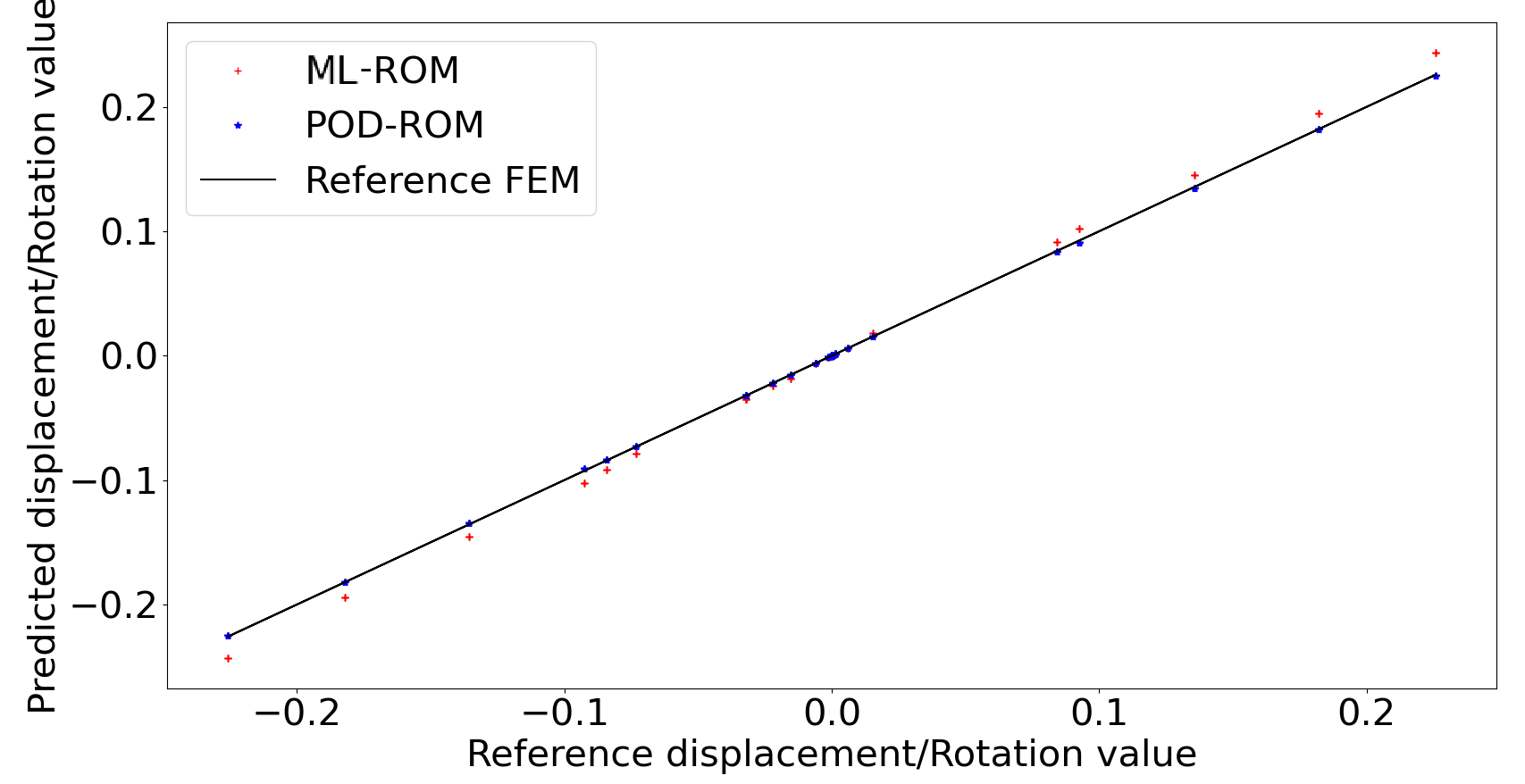}
\caption{{\em Geometrically nonlinear plate problem}: Comparison of the predicted final displacement using the Machine Learning based Reduced Order Model (ML-ROM) with that obtained from the classical Proper Orthogonal Decomposition ROM (POD-ROM) and the one from the Full Order Model (FOM), i.e., a standard finite element solver.}
\label{fig:NLGeo_Ulast}
\end{figure}

Figure~\ref{fig:NLGeo_Phi_1_5} shows the evolution of $\bm\theta_1$ and  $\bm\theta_5$ as a function of the increment step number. For $\bm\theta_1$, the first neural network output, the curve exhibits a nonlinear behavior, yet it demonstrates relatively smooth variations. The machine learning algorithm successfully reproduces the curve by generating precise predictions values at different increment steps, whether in the training or validation step. However, the curves of $\bm\theta_j$ for $j \in [1,5]$ become more complex with increasing $j$, and the prediction accuracy decreases as it clearly appears on the prediction of $\bm\theta_5$. Nevertheless, the influence of $\bm\theta_j$ on the global solution diminishes with increasing $j$. This results in the displacement prediction of Figure~\ref{fig:NLGeo_Ulast} closely matching the classical reduced order model (POD-ROM) displacements. The ML-ROM differing by less than $7.5 \%$, as a maximum, from the standard finite elements solution, which is used as a reference solution.\\

This simple geometrically nonlinear use case demonstrates that the methodology used in Section~\ref{sec:linearparam} is promising in nonlinear scenarios, such as the ones exhibiting geometrically nonlinearities. Geometrical nonlinear parametric use cases are the topic tackled in the next section.

\subsection{Displacement prediction as function of material parameters}\label{sec:plateparametric}

In this section, the same plate scenario is considered again, as in the previous section, maintaining identical loading and boundary conditions. However, in this example, we introduce variations in the Young's modulus, Poisson coefficient, and plate thickness, mimicking the setup in Section~\ref{sec:linearparam}. Essentially, we replicate the use case of Section~\ref{sec:linearparam} to encompass geometrically nonlinear behavior under higher loads.\\

A database of 65 finite element simulations spanning the same parameter range as in Section~\ref{sec:linearparam}, is constructed. Each simulation is subdivided into $N_t$ time increments, where $N_t$ remains constant across all the simulations, set to $N_t=118$. Each displacement vector is thus associated with a specific parameter combination and time step increment, forming the snapshots utilized to apply the methodology outlined in Section~\ref{sec:linearparam}. To assess the quality of the results, we designate 6 random use cases as validation instances, reserving the remaining 59 use cases for training.\\

Conducting a validation test involves supplying our software with input parameters $\mu=(E, \nu, e)$ and initiating a reduced order model simulation. This simulation utilizes the reduced displacement $\xiv_{i-1}$ from the previous time step $t_{i-1}$ to forecast the inverse of the reduced stiffness matrix at the current time step $t_{i}$. Subsequently, this enables the computation of $\xiv_{i}$ for all time steps $i \in [1, N_t]$. Unlike the plate example outlined in Section~\ref{sec:platestatic}, a test use case entails the complete execution of the time-dependent simulation associated with a specific parameter combination, circumventing the necessity of the FE solver. The geometry and boundary conditions are the same across all use cases. The determination of the reduced matrix coefficients $\underline{\bm \theta}_i$ at a particular time step $i$ depends on the preceding time step outputs, namely, the solution for reduced coordinates $\xiv_{i-1}$ in addition to the chosen time increment $\delta t_i$, and the selected parameters set $\mu$. Mathematically, this can be expressed as:
 $$\underline{\bm \theta}_i = g(\mu, \delta t_i, \xiv_{i-1}),$$
Once predicted, $\underline{\bm \theta}_i$ facilitates the deduction of the inverse of the stiffness matrix ${\mathcal{A}^{-1}_i}$ at time step $t_i$.\\

$\bm \theta_1$ closely resembles that of Figure~\ref{fig:NLGeo_Phi_1_5}. In general, $\bm \theta_k$ exhibits similar characteristics to those noted in the previous sections, with complexity escalating alongside the order of reduced space $k$. Although predictions may exhibit decreased accuracy in such scenarios, the overall impact on the global solution tends to be minimal, given the reduced influence of $\bm \theta_{k}$ on the solution with increasing $k$.\\

As for the choice of the Machine Learning algorithm, we recall that Random Forest is used across this research work for the reasons previously mentioned in section \ref{sec:linearparam}.\\

\begin{figure}
\centering
        \begin{subfigure}{0.48\textwidth}
            \includegraphics[width=\textwidth]{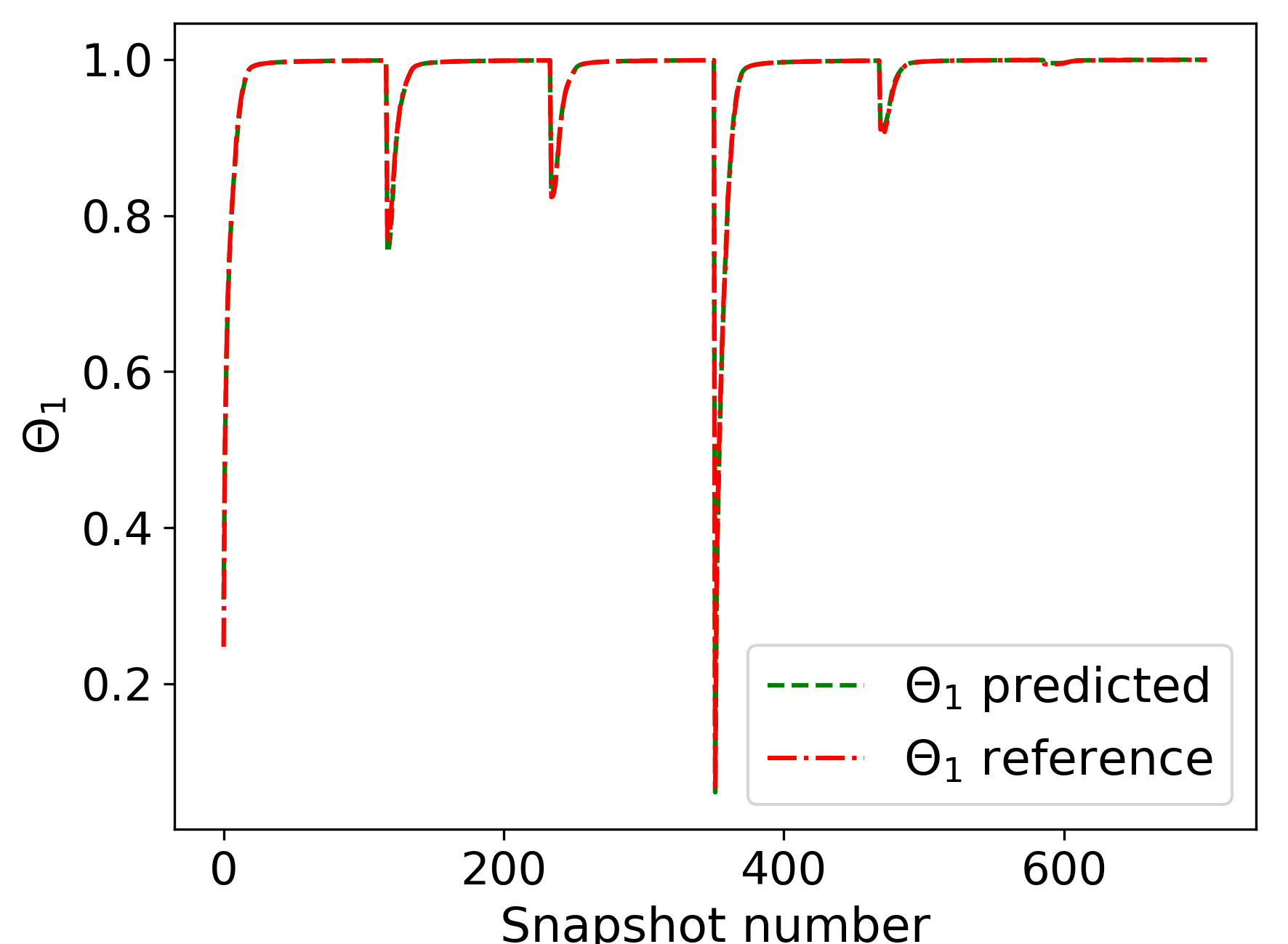}
            \caption{Validation data set results for $\bm \theta_{1}$}
            \label{fig:NLGeo_param_phi1test}
        \end{subfigure}  
    \hfill
        \begin{subfigure}{0.48\textwidth}
            \includegraphics[width=\textwidth]{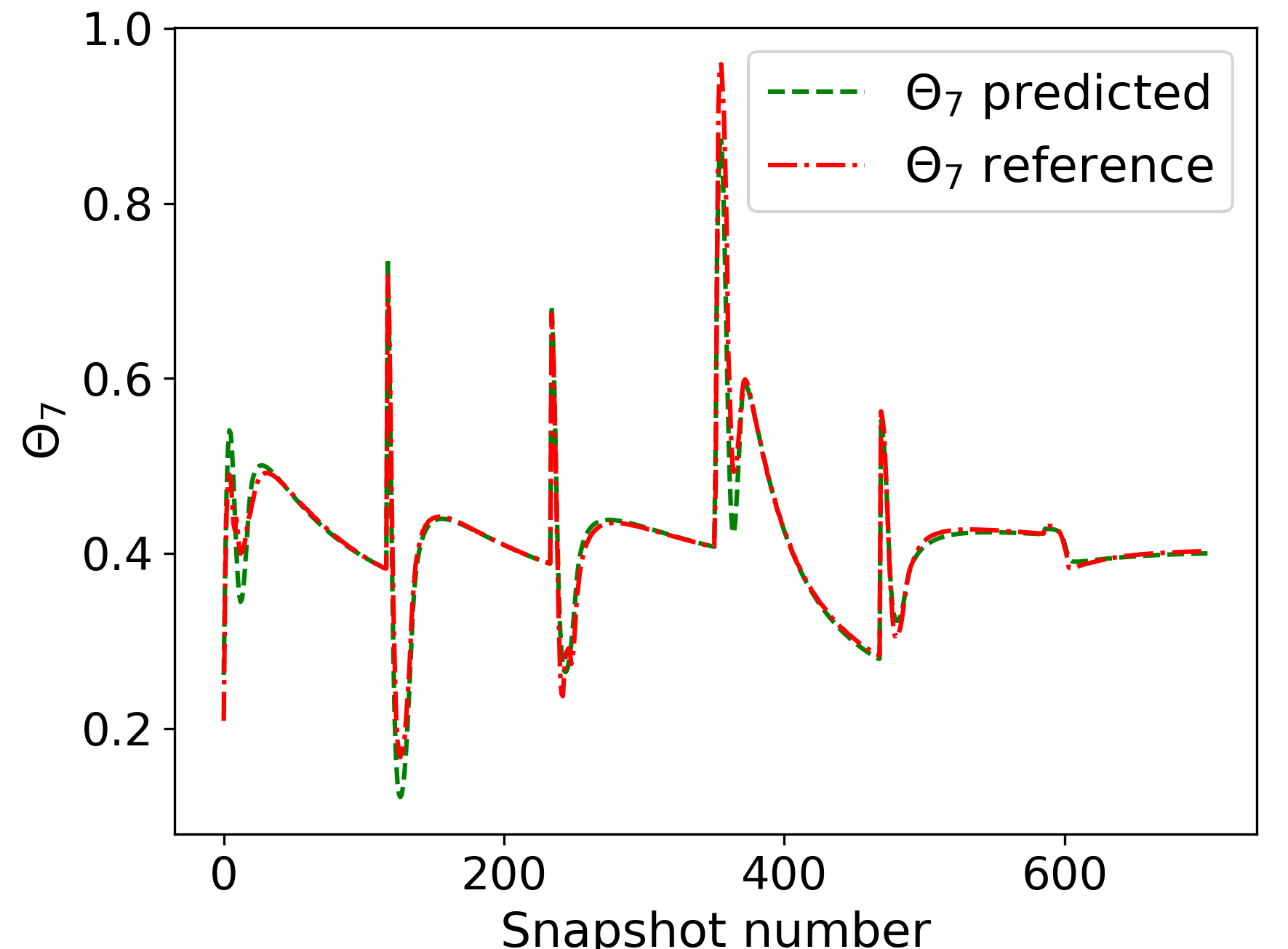}
            \caption{Validation data set results for $\bm \theta_{7}$}
            \label{fig:NLGeo_param_phi7test}
        \end{subfigure}  
\caption{{\em Geometrically nonlinear plate problem}: Comparison of reference and predicted values for $\bm \theta_{1}$ and $\bm \theta_{7}$ on the the validation data sets.}
\label{fig:NLGeo_param_phis}
\end{figure}

Figure~\ref{fig:NLGeo_param_phis} compares the predicted versus reference values for different $\bm \theta_{k}$ values for $k=1, 7$ on the validation datasets. The values are normalized in the range $[0,1]$ to facilitate machine learning training. Figure~\ref{fig:NLGeo_param_phi1test} demonstrates the efficacy of machine learning predictions on test cases in reproducing $\bm \theta_{1}$ values, despite significant differences or peaks between use cases. This indicates robust performance in capturing complex patterns. Even for $\bm \theta_{7}$, which exhibits larger variations, the prediction remains accurate, as depicted in Figure~\ref{fig:NLGeo_param_phi7test}. It's worth noting that the accuracy of the prediction is even higher on training datasets, though not explicitly shown in this discussion.\\

\begin{figure}
\centering
\includegraphics[width=\textwidth]{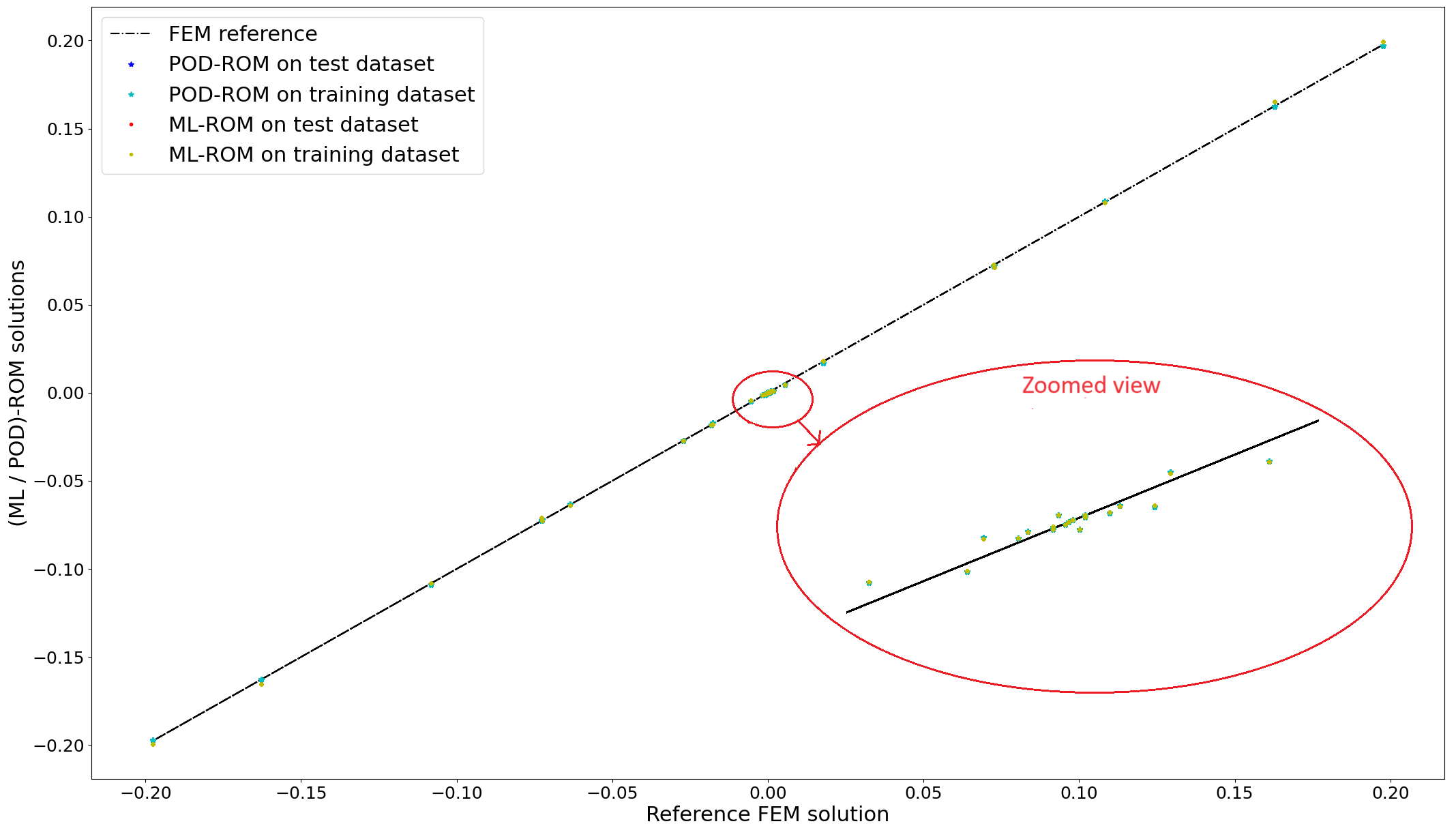} 
\caption{{\em Geometrically nonlinear plate problem}: Assessing the accuracy of the prediction of displacements and rotations for the Machine Learning based Reduced Order Model (ML-ROM) and Proper Orthogonal Decomposition based ROM (POD-ROM) compared with the finite element based Full Order Model (FOM).}
\label{fig:NLGeomParam_Uall}
\end{figure}

The distinct $\theta_{k}$ values, with $k \in [1,7]$ in this example, play a pivotal role in reconstructing the displacement solution, as elaborated in the preceding section. Remarkably, the displacement results obtained across various training or test cases closely align with those of the POD-ROM solution, which itself closely mirrors the reference FEM solution, as illustrated in Figure~\ref{fig:NLGeomParam_Uall}. When compared to the FEM reference solution, the predicted displacement using ML-ROM demonstrates minor discrepancies. Even in extreme cases, the maximum error on displacement or rotation across all test cases does not exceed $8.7 \%$. On average, the error remains notably lower, averaging at approximately $3 \%$ across all test cases.

\subsection{Conclusion}
In this section, we delved into extending the method from parametric linear elastostatic problems to parametric geometrically nonlinear static problems. This extension is illustrated using a simple plate geometry meshed with shell elements. To showcase the method's robustness on a higher complexity geometry, and higher degrees of freedom problems, Section~\ref{sec:airfoil} presents an application on an airfoil geometry subjected to a pressure load, showcasing significant deformation and rotation.

\section{Illustrating the method on a complex geometry}\label{sec:airfoil}
\subsection{Study case presentation}
\begin{figure}
\centering
    \begin{subfigure}{0.57\textwidth}
    \includegraphics[width=\textwidth]{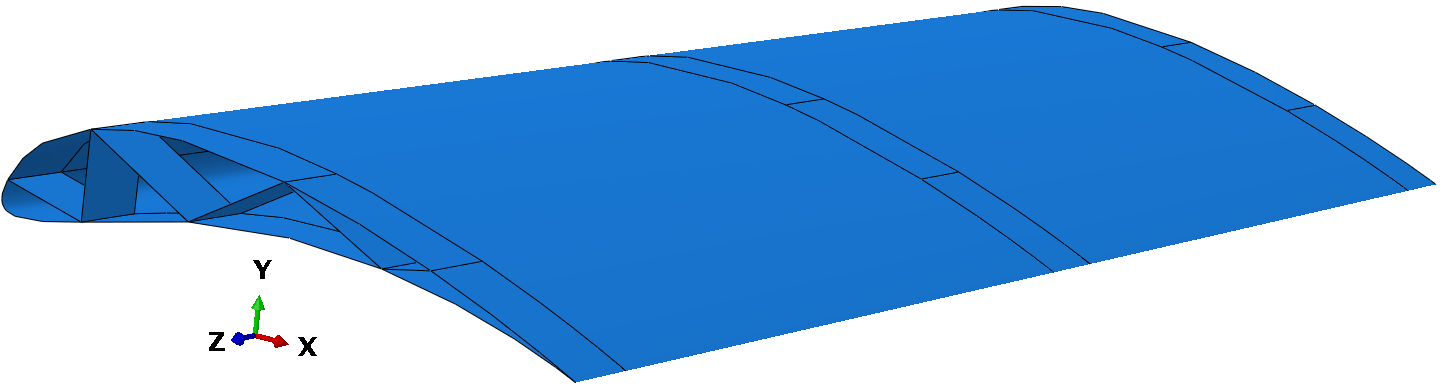}
    \caption{General view of the considered geometry.}
    \label{fig:Airfoil_geometry}
    \end{subfigure}
\hfill
    \begin{subfigure}{0.37\textwidth}
    \includegraphics[width=\textwidth]{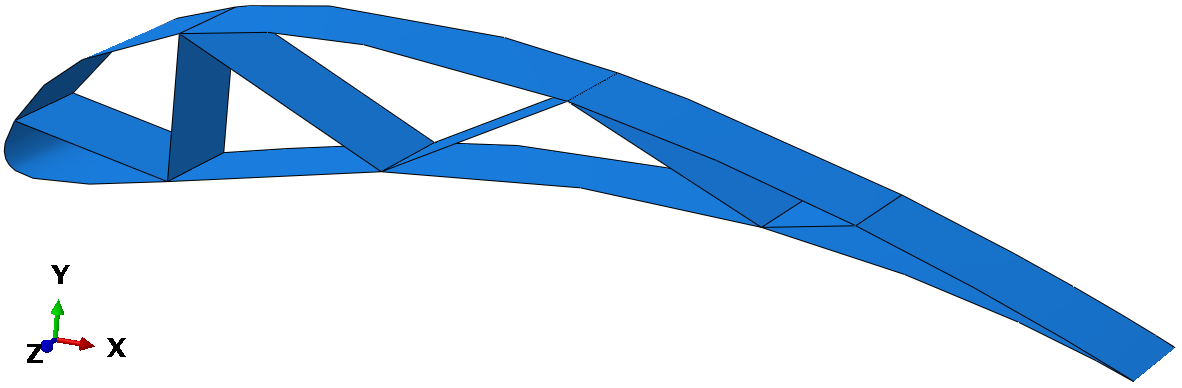}
    \caption{Geometry of reinforcement trusses.}
    \label{fig:Airfoil_trusses}
    \end{subfigure}
\caption{{\em The airfoil problem}: Illustration of the airfoil geometry used for Machine Learning based Reduced Order Model (ML-ROM).}
\label{fig:Airfoil}
\end{figure}

This section extends the methodology previously applied in Section~\ref{sec:NLGEO} to a more complex structure, aiming to test its robustness and readiness for industrial applications beyond simple academic problems.  Figure~\ref{fig:Airfoil} depicts an airfoil structure comprised of shell elements, reinforced by three trusses, one at the middle of the airfoil, and one at each side. These trusses' upper surfaces coincide with the airfoil surface and are welded to the airfoil. In this example, we consider three parameters: a Young's Modulus shared by the trusses and the airfoil, ranging from 120 GPa to 300 GPa; the external surface thickness of the structure, denoted as $e_1 \in [1, 5] $ mm; and the truss beams' thickness, denoted as $e_2 \in [5, 7.5] $ mm.\\

The airfoil structure is fixed at the left edge and free on the right one. A uniform surface pressure $P= 2.7 MPa$ is applied to the lower surface of the airfoil. The structure exhibits nonlinear geometrical behavior while its material is linear elastic. The structure is meshed with linear shell elements providing 6300 degrees of freedom on the 1050 nodes, a considerable increase compared to the examples previously illustrated in earlier sections of this work.\\

To cover the parameters' range, a total of 100 calculations are performed, with 8 covering the parameter limits and the remaining 92 selected using the Latin Hypercube method within the parameter ranges. Abaqus is utilized to execute these simulations. For each simulation, the displacement vector is extracted for every time increment and added to the snapshot database. Similarly, the load vector and the stiffness matrix are recorded. The reader is refered to appendix \ref{Abaqus} for more details on the technical implementation.\\

For the training process, $85\%$ of the snapshots will be used as a training set, while the remaining $15 \%$ will serve as a validation dataset. The validation dataset is randomly chosen from the available set of simulations. However, parameter combinations corresponding to the parameter limits are always included in the training dataset, to prevent extrapolation when predicting on unseen parameter combinations. The methodology is the same as already exposed in Section~\ref{sec:NLGEO} leading to $\bm \Theta$ matrix of size $(\mathtt P,9)$, where $\mathtt P$ is the number of available snapshots. The values of $\bm \Theta$  are preprocessed using the Z-score method \cite{Zscore} to accelerate the training process. 

\subsection{Results discussion}
As illustrated in previous examples, the accuracy of the results in this study hinges on the precision of predicting $\bm \Theta$. A precise prediction ensures that the ML-ROM aligns with the POD-ROM solution, which, for this case and all snapshots, exhibits negligible discrepancies from the FEM solution.\\

In this example, Random Forest is used for predicting the values of $\bm \Theta$. The Grid Search Cross Validation methodology (GridSearch) \cite{GridSearch} is used to optimize the Random Forest parameters. The chosen configuration includes 50 estimators, 4 minimum sample leaves, 70 random states, and the use of bootstrap. In fact, for complex study cases, Grid Search can act as an automation tool to efficiently select the optimal Random Forest parameters, reducing the reliance on user trial and error.\\

For simplicity, we refrain from showcasing all $\bm \theta_k$ values for $k \in[1,9]$. Instead, we focus primarily on $\bm \theta_1$, which holds the most significant impact on results, and $\bm \theta_9$, which has the least impact and is generally is the most challenging to predict accurately.\\

\begin{figure}
\centering
    \includegraphics[width=\textwidth]{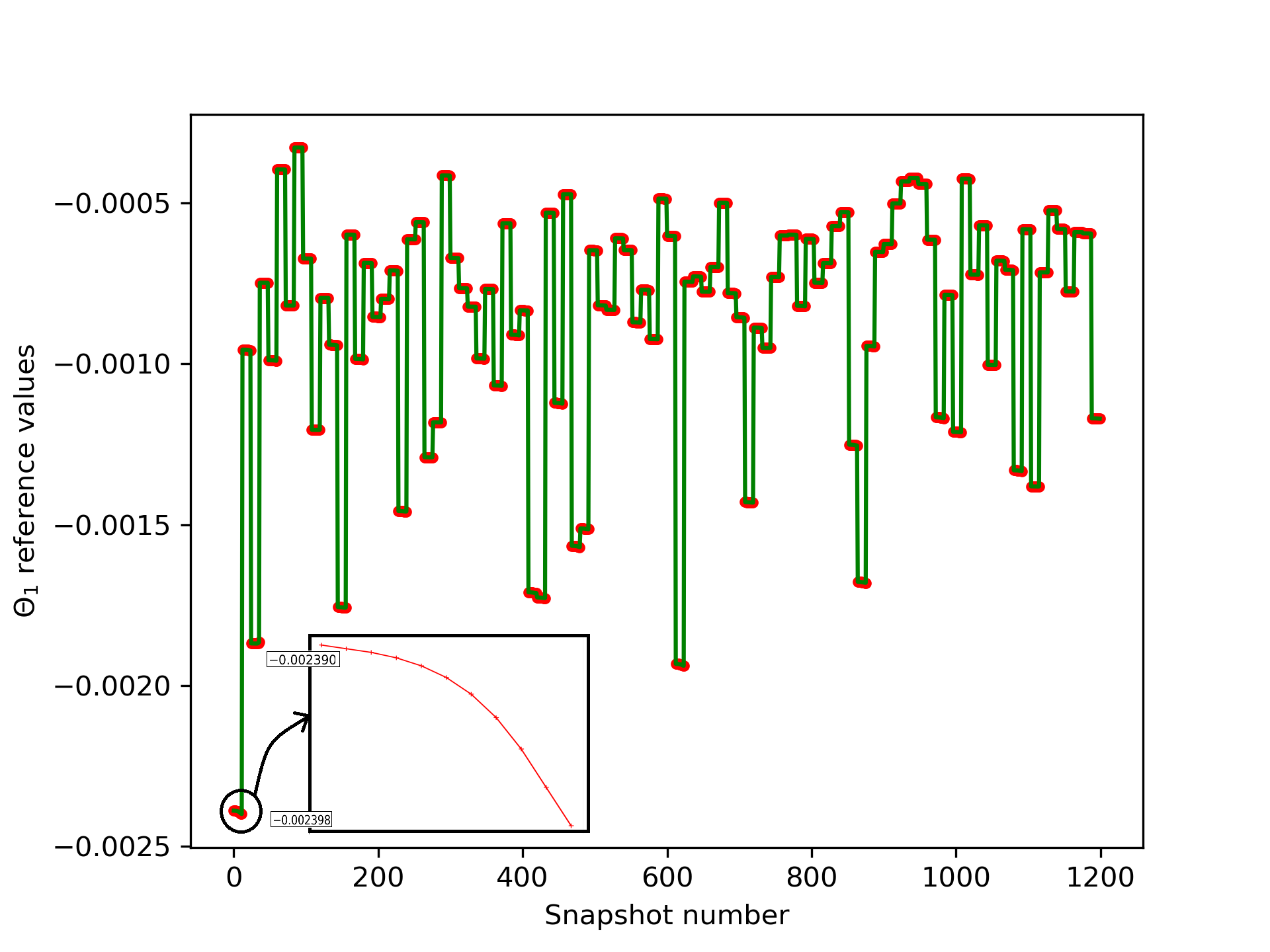}
    \caption{{\em The airfoil problem}: $\bm \theta_{1\text{ref}}$ as function of the snapshot number.}
    \label{fig:Airfoil_Theta1_allVal}
\end{figure}

To begin with, the reference values of $\bm \theta_1$, denoted as $\bm \theta_{1\text{ref}}$, are visually represented in Figure~\ref{fig:Airfoil_Theta1_allVal}. Each of the 100 selected study cases encompasses multiple time increments, yielding a total of $N_t$ snapshots. Examining Figure~\ref{fig:Airfoil_Theta1_allVal}, it becomes apparent that within a singular study (depicted in red), the evolution of $\bm \theta_1$ is characterized by minimal variations. However, traversing from one parameter set to another, represented by the green curve connecting all snapshots, reveals significant alterations. It's crucial to note that while connecting outcomes from distinct study cases lacks physical validity, this approach effectively emphasizes the transitions in $\bm \theta_{1\text{ref}}$ across different study cases. This observation underscores a primary challenge for the machine learning algorithm: accurately capturing both the values associated with a particular use case and their temporal evolution within said use case.\\

Similar conclusions can be drawn for $\bm \theta_{k}$ with $k \in [2,9]$. However, as $k$ increased, $\bm \theta_k$  exhibits higher fluctuations and lower values compared to $\bm \theta_1$, posing additional challenges for accurate predictions using machine learning.\\ 

\begin{figure}
\centering
    \includegraphics[width=\textwidth]{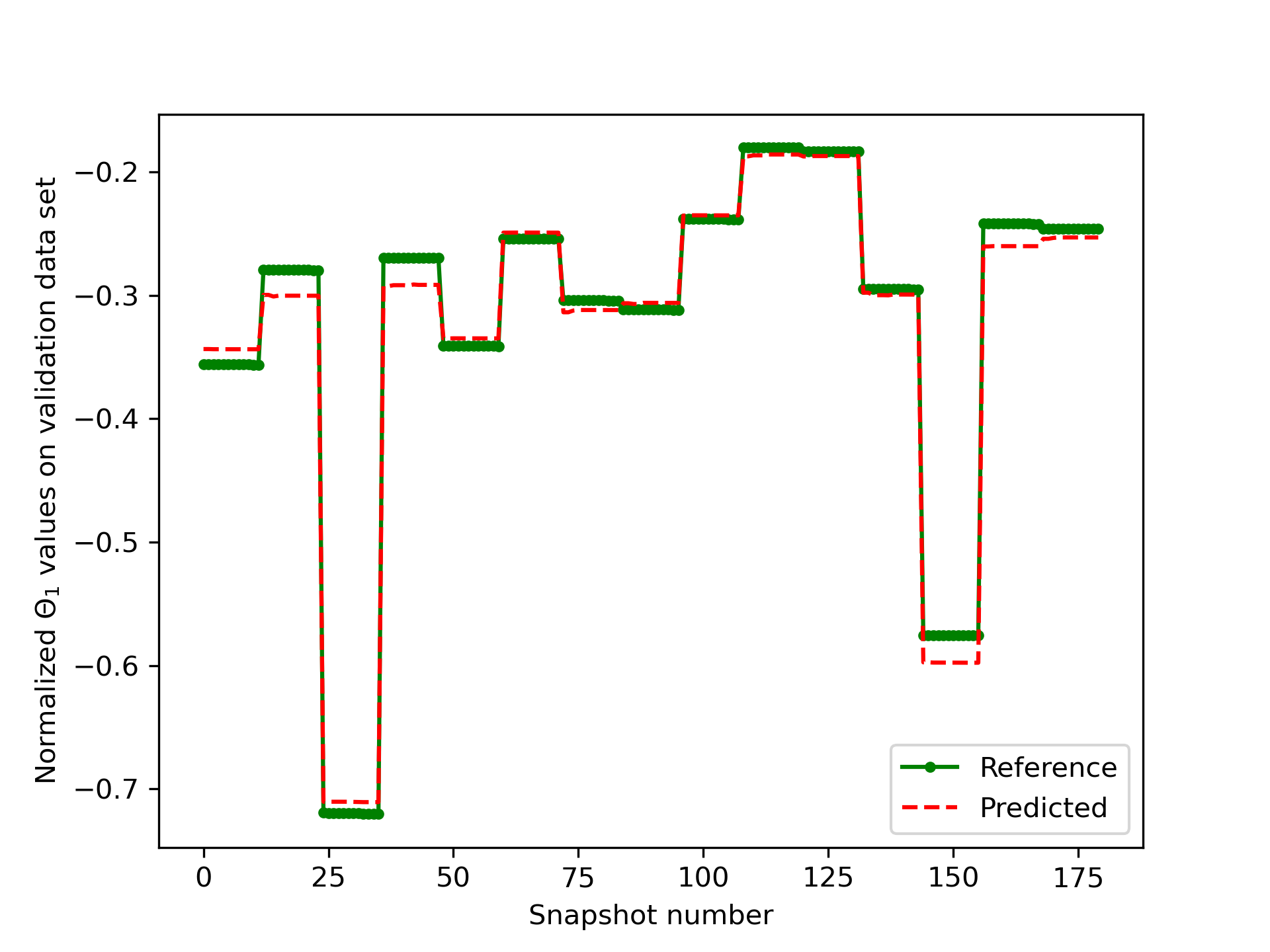}
    \caption{{\em The airfoil problem}: Predicted versus reference $\bm \theta_{1}$ values for all snapshots of the validation data set.}
    \label{fig:Airfoil_Theta1_test}
\end{figure}
Figure~\ref{fig:Airfoil_Theta1_test} showcases the prediction results obtained from the validation dataset for $\bm{\theta}_{1}$. Notably, the values presented in this figure are those associated with Z-score normalization. The machine learning model demonstrates commendable performance in reproducing results from the validation dataset, albeit with slightly higher accuracy observed in the training dataset. This trend persists across the other $\bm{\theta}_k$ values for $k \in [2,9]$.\\

Detailed analysis of the prediction on $\bm{\theta}_k$ values is not the primary focus. It is worth noting that the accuracy of displacement values is of greater interest. Displacement errors exhibit negligible deviations on the training dataset, with errors consistently below $0.25\%$ across all degrees of freedom. In fact, the differences between the predicted values and those of the reference POD-ROM are imperceptible. However, on the 15 validation use cases, the average relative error in displacements is $0.7\%$, although it peaks at $7\%$ for a single parameters combination case.

\begin{figure}
\centering
    \includegraphics[width=\textwidth]{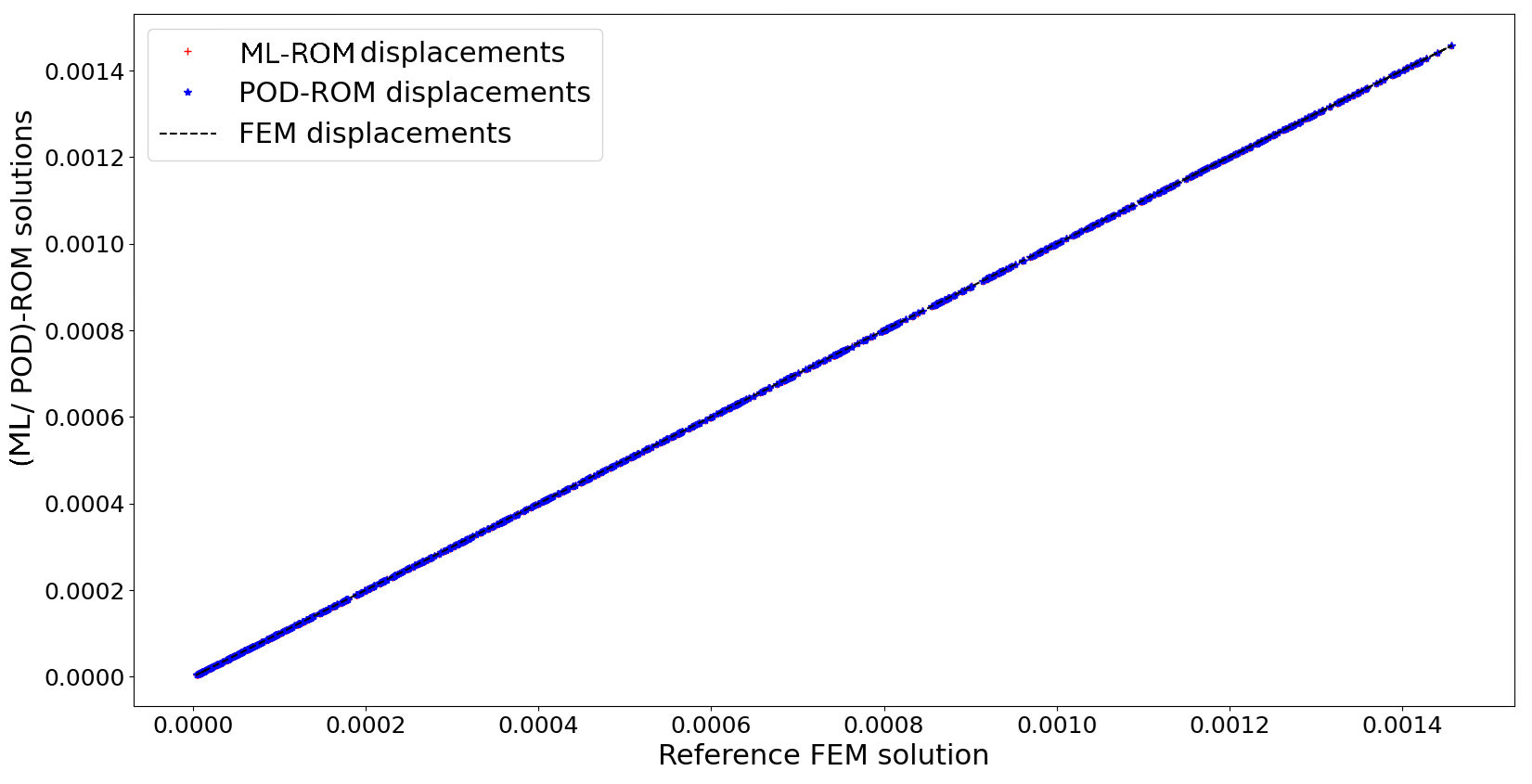}
    \caption{{\em The airfoil problem}: Vertical displacement comparison on a validation data set exhibiting low prediction error.}
    \label{fig:Airfoil_U3_case76}
\end{figure}

On the validation data set, Figure~\ref{fig:Airfoil_U3_case76} illustrates a selected validation case. We can see that the ML-ROM solution is practically identical to the POD-ROM solutions. This translates into negligible displacements error not exceeding 0.035 $\%$. 

\begin{figure}
\centering
    \includegraphics[width=\textwidth]{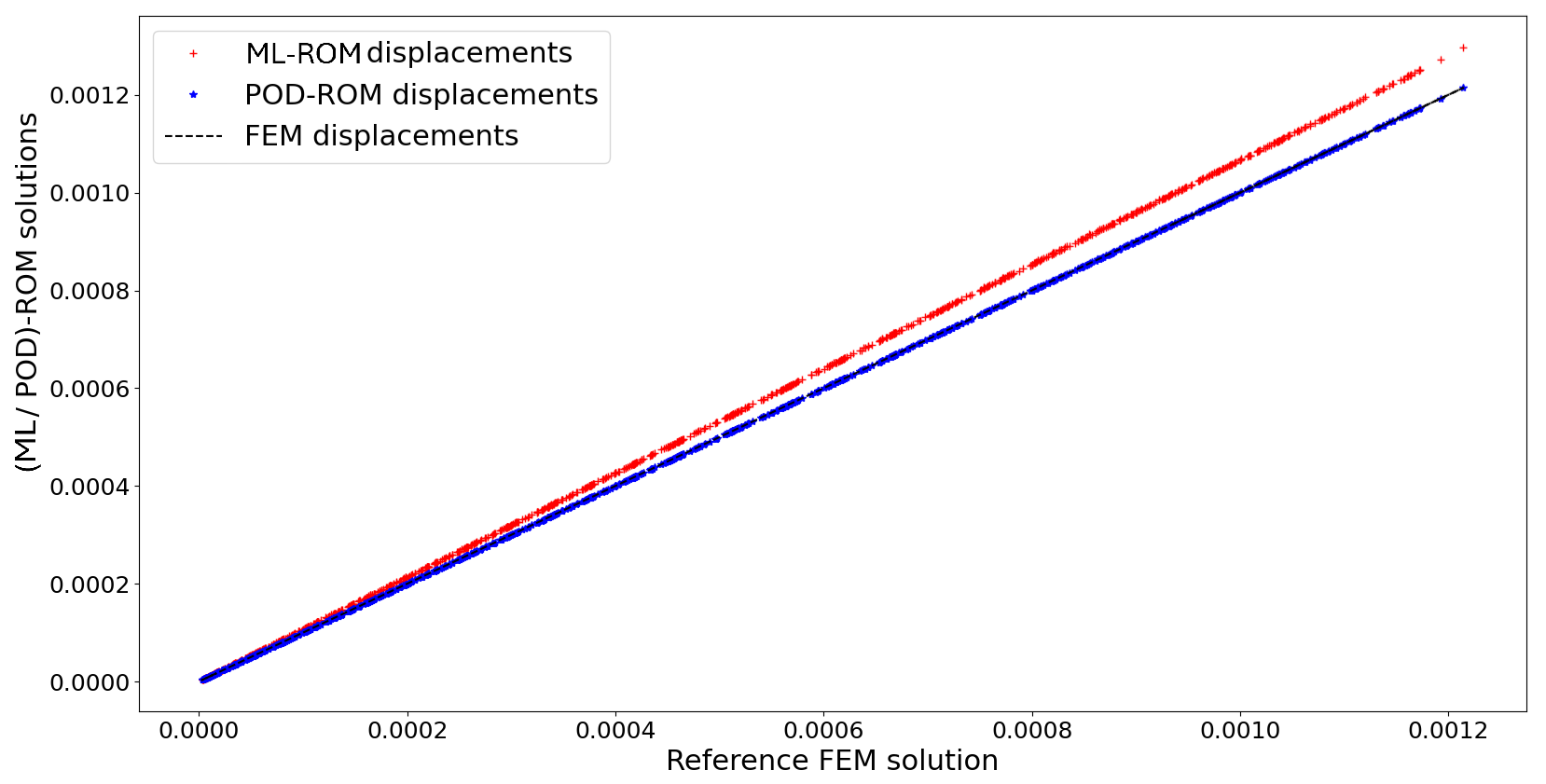}
    \caption{{\em The airfoil problem}: Vertical displacement comparison on a validation data set exhibiting the highest prediction error.}
    \label{fig:Airfoil_U3_case96}
\end{figure}

On the other hand, the use case that exhibits the highest displacement error is shown in Figure~\ref{fig:Airfoil_U3_case96}.  It is obvious that the predicted deflection deviates for this extreme case from the reference one, however, this deviation remains acceptable a relative error not exceeding $7 \%$. 

\subsection{Conclusion}
The analysis conducted using the airfoil examples demonstrates the method's robustness in handling complex geometries and loading conditions. Remarkably, the method's effectiveness and accuracy remained uncompromised despite encountering increased geometric complexity, and even though predicting $\mathbf \Theta$ seemed to present greater challenges.
\section{Conclusions and perspectives}\label{finalsonclusionssec}
In this article we have developed a lightly intrusive reduced order model empowered by machine learning techniques to address affine and non-affine parametric linear and non-linear solid and structural problems. 

The present approach inherit the physical properties of the Full Order Model (FOM) in a lightly intrusive manner by requiring access to the coefficient matrix, right hand side, and the Dirichlet boundary conditions which is typically enabled by commercial software.

The novelty in the present method is the use of Proper Orthogonal Decomposition (POD) in two steps and then using machine learning to compute the inverse of the reduced order stiffness matrices for any choice of the problem's parameters.

The efficiency of the method was validated through both linear and geometrical non-linear structural academic and complex geometry problems, affirming its capability to construct real-time solutions accurately. Central to its success is the precise prediction of a reduced inverse stiffness matrices, facilitating real-time simulations without reliance on standard FEM software for matrix reconstruction. This is particularly significant in contexts involving nonlinear behavior, where generating the tangent stiffness matrix during Newton-Raphson iterations incurs substantial computational costs. By eliminating this dependency, the method enables efficient parametric studies even in scenarios characterized by computationally expensive nonlinearities.

Through ongoing development and refinement, this lightly intrusive reduced order model promises to be a valuable tool for efficiently tackling a wide array of complex parametric nonlinear and non-affine parametric problems.

\backmatter

%
%
%

\bmhead{Acknowledgements}
The authors acknowledge the support of ESI Group, RTE and SKF through their research chairs at Arts et Metiers Institute of Technology (ENSAM), and the support of Fondation Arts et Metiers during the sabbatical stay at ENSAM for the fourth author

\section*{Declarations}
The authors declare the following:
\begin{itemize}
\item Funding: this project has received funding from the European Defence Fund (EDF) under grant agreement 101103257  dTHOR, EDF-2021-NAVAL-R-2.
\item Conflict of interest: the authors declare no conflict of interest/competing interests
\item Data availability: there are no specific data related to this publication. 
\item Code availability: the first author could make some codes available on the request of the readers.
\item Author contribution: the second and last authors primarily contributed to the conceptualization of the method. The first author was responsible for developing the methodology, conducting simulations, training the machine learning models, and drafting the initial manuscript. All authors contributed to the validation of results and provided critical revisions to the manuscript. The second and last authors also played an active role in supervising the overall work and ensuring its scientific rigor.
\end{itemize}

\begin{appendices}

\section{PGD-based regularized regression}\label{pgdregression}
Without loss of generality, we describe the approximation of a scalar function $f$ that depends on the pair of parameters $(\mu,\eta)$. We assume $f^i \equiv f(\mu_i,\eta_i)$ are known at the $\mathtt P$ points of the sampling campaign.\\

Here, the scalar $f$ represents each component $\mathbb S_{pq}$, $p=1, \ldots, \mathtt K$ and $q = 1, \ldots , \mathtt L$. The goal is to find a finite sum decomposition (PGD) approximation of function $f(\mu,\eta)$, expressible as a finite sum of $\mathtt M$ terms, according to
$$
f(\mu,\eta) \approx  f^M(\mu,\eta)=\sum_{j=1}^\mathtt M \mathcal G_j(\mu) \mathcal H_j(\eta),
$$
able to approximate the known data $f^i \equiv f(\mu_i,\eta_i)$, $i=1, \ldots, \mathtt P$.\\

Note that in two-dimensional scenarios we do not need, in general, to construct separated representations. However, our aim here is to introduce the technique for a general enough problem, eventually involving numerous parameters.\\

We propose a greedy algorithm that works by looking, at iteration $m$, for the update $\mathcal G_m(\mu) \mathcal H_m(\eta)$ such that $f_m(\mu,\eta) = f_{m-1}(\mu,\eta) + \mathcal G_m(\mu) \mathcal H_m(\eta)$. To compute the sought functions,  we first approximate them as $\mathcal G_m(\mu) = \mathbf N_m^{\mu,T}(\mu) \bs a_m$ and $\mathcal H_m(\eta) = \mathbf N_m^{\eta,T}(\eta) \bs b_m$, where $\mathbf N_m^\mu$ represents the  basis considered for approximating the $m$-th mode, that depends on the $\mu$-parameter, with $\bs a_m$ the associated weights. We proceed similarly for the other direction ($\eta$-parameter): $\mathbf N_m^\eta$ and $\bs b_m$.

The updated solution is found by solving the minimization problem:
$$
\mathcal G_m(\mu) \mathcal H_m(\eta) = \argmin_{(\mathcal G_m(\mu) \mathcal H_m(\eta) )^\ast}
\sum \limits_{i=1}^{\mathtt P} \| f^i - \left(f^i_{m-1}+ (\mathcal G_m(\mu_i) \mathcal H_m(\eta_i))^\ast\right) \|_2^2,
$$
that, by using the notation below, reads
\begin{align*}
\bs{r}
&
=
\begin{pmatrix} 
f^1-f_{m-1}(\mu_1,\eta_1) 
\\ 
\vdots 
\\
f^{\mathtt P}-f_{m-1}(\mu_{\mathtt P},\eta_{\mathtt P}) 
\end{pmatrix},
\\
\mathbb{M}_\mu
&
=
\begin{pmatrix} 
\mathbf N_m^{\eta,T}(\eta_1) \bs{b}_m 
\mathbf N_m^{\mu, T}(\mu_1)) 
\\ 
\vdots 
\\
\mathbf N_m^{\eta,T}(\eta_{\mathtt P}) \bs{b}_m 
\mathbf N_m^{\mu, T}(\mu_{\mathtt P})) 
\end{pmatrix},
\\
\mathbb{M}_\eta
&
=
\begin{pmatrix} 
\mathbf N_m^{\mu,T}(\mu_1) \bs{a}_m 
\mathbf N_m^{\eta, T}(\eta_1)) 
\\ 
\vdots 
\\
\mathbf N_m^{\mu,T}(\mu_{\mathtt P}) \bs{a}_m 
\mathbf N_m^{\eta, T}(\eta_{\mathtt P})) 
\end{pmatrix},
\end{align*}
results in the two problems:
\begin{equation}
\label{ridge2x}
 \bs{a}_m = \argmin_{\bs{a}_m^\ast}
\Big\{
\|\bs{r} - \mathbb{M}_\mu \bs{a}_m^\ast \|_2^2
\Big\},
\end{equation}
\begin{equation}
\label{ridge2y}
 \bs{b}_m = \argmin_{\bs{b}_m^\ast}
\Big\{
\|\bs{r} - \mathbb{M}_\eta \bs{b}_m^\ast \|_2^2
\Big\}.
\end{equation}
This is solved iteratively until a fixed point, whose existence is assumed, is found. 

This algorithm is at the heart of the sparse-PGD (sPGD) constructor \cite{spgd1}. However, overfitting appears when combining rich approximation bases $\mathbf N_m^\mu(\mu)$ and $\mathbf N_m^\eta(\eta)$, with low sampled data-sets $\mathtt P$. Such a situation is always the case when operating in highly multi-parametric settings.\\

To avoid overfitting, an adaptive procedure was proposed in \cite{spgd1} that consists in adapting the approximation basis, such that their degree increases when advancing in the modal enrichment $m$. Looking for a more versatile and automatic method, different regularizations were also proposed in \cite{SAN21}.\\

When the data set is rich enough, that is $\mathtt P \gg 1$, a nonlinear general regression $f(\mu,\eta)$ could be constructed by making use of a machine learning algorithm, $f^{ML} (\mu,\eta)$.

\section{Implementation of the Proposed Method on Commercial FEA Software}\label{Abaqus}
This appendix aims to provide a technical guide on implementing the proposed method in a commercial Finite Element Analysis (FEA) software. While the explanation is intended to be general and independent of the specific software chosen, certain details are tied to the software utilized in this study, namely Abaqus.

The method described in this paper is divided into two phases: an offline (training) phase and an online phase. The offline phase requires a Full-Order Model (FOM) solver, while the online phase operates with reduced matrices that are lightweight and can be efficiently handled in Python or other programming languages. Automation of the training phase is possible by developing a Python script or an Abaqus plugin to execute the Design of Experiment (DOE), collect snapshots (including stiffness matrices), apply Proper Orthogonal Decomposition (POD), and use GridSearch with Random Forest to create a machine learning model that predicts the coefficients needed to construct the stiffness matrix for a given set of material parameters. In the online phase, interaction with the FOM solver is unnecessary, although it remains an option if required.

To apply the method described in this paper, the FEA software should provide four main outputs:
\begin{itemize}
	\item \textbf{Solution fields}
    \item \textbf{The stiffness matrix}: For non-linear applications where the stiffness matrix evolves over time, the software must output this matrix at each integration step.
    \item \textbf{The load vector}
    \item \textbf{Boundary conditions}
    
\end{itemize}

This method is generalizable, allowing for the creation of snapshots, the application of the method, and the development of a digital twin in a fully automated manner. 

\subsection{Extracting the solutions fields}

The user should first set up a base use case with a given set of parameter combinations. Then, they can modify the Abaqus input file so that each varying parameter (such as Young's Modulus, thickness, or Poisson's ratio) is read from an external input file, which is managed by a Python script. This script selects parameter values according to the DOE, updates the input files, and launches the Abaqus solver. After the solver completes the calculations, the script opens the \texttt{.odb} file (Abaqus' results file in binary format) using the appropriate Abaqus-Python Library (\texttt{Visualization}), reads the displacement results, and records them in an HDF5 file where the snapshots are stored. Before proceeding to the next calculation, the script should also read the stiffness matrix. Here are the challenges involved and how they can be addressed:

\subsection{Extracting the Stiffness Matrix in Abaqus}

There are two main scenarios for extracting the stiffness matrix:

\begin{itemize}
    \item \textbf{Constant Stiffness Matrix:} If the stiffness matrix is constant for the given calculation, it can be extracted at the beginning of the simulation. This is easily done by creating a dedicated step in the input file to generate the stiffness matrix, which is output as a \texttt{.mtx} file in sparse format. This file can be read using the \texttt{scipy} library in Python, specifically using the sparse functionality. Generating and reading such a matrix is straightforward and fast. However, the matrix will not have imposed boundary conditions, so the user must modify it, such as by removing rows and columns corresponding to fixed boundary conditions.
    
    \item \textbf{Evolving Stiffness Matrix:} For nonlinear problems where the stiffness matrix evolves during the calculation, the matrix is needed at each iteration step. Abaqus does not directly provide an assembled matrix but outputs a \texttt{.mtx} file with the elemental matrices at each increment. The user must write a Python script to parse the \texttt{.mtx} files and assemble the matrices for each step increment using the connectivity matrix that can be obtained by reading the input file using a Python script. While this adds complexity, it is automatable since the file format is standardized. Despite being computationally demanding, extracting the matrix with Python is often less intensive than running the simulation itself.
\end{itemize}

\subsection{Additional Requirements: Load Vector and Boundary Conditions}

Once the stiffness matrix and solution field vectors are obtained, the user must also retrieve the boundary conditions and load vectors. If the load is known (e.g., a point force applied to a specific node), this information can be directly written in vector form. For more complex loading conditions, there are two approaches:

\begin{itemize}
    \item \textbf{Point Load:} Point loads can be easily read from the \texttt{.odb} file, similar to accessing displacements or velocities.
    
    \item \textbf{Distributed Load (e.g., Pressure):} For distributed loads, like the example in Section~\ref{sec:airfoil}, the best approach is to output this vector to a dedicated \texttt{.mtx} file, similar to the stiffness matrix. If the load is constant, no assembly is required, and it can be extracted at the beginning of the simulation.
\end{itemize}

Boundary conditions, which are defined in the input file, can be easily read and translated into mathematical constraints using Python. In our study, fixed boundary conditions were imposed on known node points, which were straightforwardly converted into a Python array for later use.

\section{Nomenclature and notation}
{
\renewcommand{\arraystretch}{1.3}
\begin{center}
\begin{tabular}{lp{.7\textwidth}}
$\mathcal{P}$ & {\em Parameter space}: Set of valid parameter choices to the problem \\\hline
$\bm \mu \in \mathcal{P}$ & {\em Parameter instance}: A particular instantiation of parameters \\\hline
$\mathtt{P}$ & Number of samples available for POD \\\hline
$N$ & Number of degrees of freedom in the full order model \\\hline
$n$ & Dimension of reduced model space \\\hline
$\xi_i$ & $i$-th reduced model degree of freedom \\\hline
$\xiv$ & Vector of the reduced model degrees of freedom \\\hline
$\lambda_i$ & Energy content of $i$-th reduced model degree of freedom \\\hline
$\bm V$ & {\em Projection matrix}: $N \times n$ matrix used to convert reduced solution vectors to high-fidelity space solution vectors. \\\hline
$\Phiv$ & Common projection matrix (for the second round of POD). \\\hline
$\bm A(\bm \mu)$ & High-fidelity system matrix \\\hline
$\bm f(\bm \mu)$ & High-fidelity load vector \\\hline
$\bm A_\text{r}(\bm \mu) = \bm V^{\intercal} \bm A(\bm \mu) \bm V$ & Reduced system matrix \\\hline
$\mathcal A(\bm \mu) = \mathtt{Vect}(\bm A_\text{r}(\bm \mu))$ & Vectorized reduced system matrix \\\hline
$\mathcal A^{-1}(\bm \mu) = \mathtt{Vect}(\bm A^{-1}_\text{r}(\bm \mu))$ & Vectorized inverse reduced system matrix \\\hline
$\bm u_{h,\text{r}} = \bm V \bm \xiv$ & Reconstructed high-fidelity solution coefficient vector \\\hline
$N_t$ & Number of time increment in a solution \\\hline
$\bm\Theta$ & Matrix of reduced coordinates representing $\mathcal A^{-1}(\bm \mu)$ in their reduced space \\\hline
$\bm\theta_k$ & Line $k$ of $\bm\Theta$ \\\hline
$\underline{\bm \theta}_i$ & Vector containing the values of $\bm\Theta$ at time step $i$ \\\hline
\end{tabular}
\end{center}
}

\end{appendices}

\clearpage


\end{document}